\documentclass[12pt,preprint]{aastex}

\slugcomment{to be appeared in Astronomical Journal}

\shorttitle{NGC 2264}
\shortauthors{Sung \& Bessell}

\begin{document}

\title{The Initial Mass Function and Young Brown Dwarf Candidates in NGC 2264.\\
IV. The Initial Mass Function and Star Formation History}

\author{Hwankyung Sung}
\affil{Department of Astronomy and Space Science, Sejong University,
    98, Kunja-dong, Kwangjin-gu, Seoul 143-747, Korea}
\email{sungh@sejong.ac.kr}

\and

\author{Michael S. Bessell}
\affil{Research School of Astronomy and Astrophysics, Australian
National University, MSO, Cotter Road, Weston, ACT 2611, Australia}
\email{bessell@mso.anu.edu.au}

\begin{abstract}
We have studied the star formation history and the initial mass function (IMF)
using the age and mass derived from spectral energy distribution (SED) fitting 
and from color-magnitude diagrams. We also examined the physical and structural
parameters of more than 1,000 pre-main sequence stars in NGC 2264 using 
the on-line SED fitting tool (SED fitter) of Robitaille et al. 

We have compared the physical parameters of central stars from SED fitter
and other methods. The temperature of the central star is, in many cases,
much higher than that expected from its spectral type. The mass and
age from SED fitter are not well matched with those from color-magnitude
diagrams. We have made some suggestions to improve the accuracy of
temperature estimates in SED fitter. 

In most cases these parameters of individual stars from SED fitter
in a star forming region (SFR) or in the whole cluster showed nearly
no systematic variation with age or with any other relevant parameter. 
On the other hand, the median properties of stars in NGC 2264 SFRs showed 
an evident evolutionary effect and were interrelated to each other. 
Such differences are caused by a larger age spread within an SFR than 
between them.

The cumulative distribution of stellar ages showed a distinct difference 
among SFRs. A Kolmogorov-Smirnov test gave a very low probability of them 
being from 
the same population. The results indicate that star formation in NGC 2264
started at the surface region (Halo and Field regions) about 6 -- 7 Myr ago, 
propagated into the molecular cloud and finally triggered the recent star 
formation in the Spokes cluster. The kind of sequential star formation that 
started in the low-density surface region (Halo and Field regions) implies 
that star formation in NGC 2264 was triggered by an external source.

The IMF of NGC 2264 was determined in two different ways. One method used
the stellar mass from the SED fitting tool, the other used the stellar mass
from color-magnitude diagrams. The first IMF showed a distinct peak at
m $\approx$ 2M$_\odot$, but the second did not. We attribute the peak 
as an artifact of the SED fitting tool because there is no such IMF with a peak 
at m $\approx$ 2M$_\odot$. The slope of the IMF of NGC 2264 for massive 
stars ($\log m \geq 0.5$) is -1.7 $\pm$ 0.1, which is somewhat steeper 
than the so-called standard ``Salpeter-Kroupa'' IMF. We also present 
data for 79 young brown dwarf candidates.

\end{abstract}

\keywords{open clusters and associations: individual (NGC 2264) -- stars:
luminosity function, mass function}

\section{INTRODUCTION}

The initial mass function (IMF) is one of the core topics in stellar astronomical 
research. The universality of the IMF has been debated for a long time (see 
\citet{sca05} for example).
In many cases the slope of the IMF of open clusters is well consistent
with the so-called standard ``Salpeter-Kroupa'' IMF within the observational
uncertainties, eg. $\Gamma = -1.2 \pm 0.1$ for
NGC 6231 \citep{sbl98}, Tr 14 and Tr 16 in the Carina nebula \citep{hsb10},
or the Arches cluster in the Galactic center \citep{esm09}.
But in some cases, a top-heavy IMF is reported for massive young open clusters
such as 
Westerlund 1 ($\Gamma = -0.8 \pm 0.1$ - \citet{clsb10}) or for young stars 
in the Galactic center stellar disk \citep{hb10}. On the other hand, a deficit 
of massive stars has also been reported for Taurus \citep{kll03} and the Cone nebula
region of NGC 2264 ($\Gamma = -1.7 \pm 0.3$ - \citet{psbk00}). In chemical
evolution models of bulges of our Galaxy and M31, a non-standard IMF has been  
favored \citep{bkm07}; however, recent precise abundance analyses of bulge and 
disk stars favors a universal IMF (G. Gilmore - private communication). The 
universality of the IMF remains an intriguing  problem to be solved in the 
near future. 
Another outstanding issue in IMF research concerns
the companion mass function in binary systems \citep{cjl06,mh09}.

The origin of the IMF is a hot topic of research.
The similarity between the dense core clump mass function and the stellar IMF
has led many investigators to implicitly assume that the clump
mass function is the origin of the stellar mass function \citep{man98,all07}.
But \citet{scb09} show the non-existence of such a correlation between clump 
mass and stellar mass. Owing to the rapid development of computing power
and techniques many attempts have been made to understand the origin
of the IMF through numerical modeling. Nowadays various physical processes such as
gravitational fragmentation \citep{hz84}, turbulence \citep{pn02}, accretion 
\citep{bbcp01}, coalescence \citep{es03}, and magnetic fields \citep{sla04}
are taken into account in the simulations. Recently \citet{sd10} were able to
reproduce IMF or clump mass functions nearly identical to observations.

Star formation history remains one of the most interesting topics of 
current research. 
It is also related to triggering mechanisms of star formation. 
Most Milky Way OB associations extend over a few degrees (linear 
size of about 100 pc). It is very difficult to imagine that star formation 
on such a large scale occurred at a single instant. We can therefore 
expect to see age differences within an OB association. Classical examples
of this are the sequential star formation in the OB associations -
Sco-Cen association (= Upper Scorpius + Upper Centaurus and Lupus + Lower 
Centaurus and Crux) \citep{pz99}, Ori OB1 \citep{ab64}
or the W3-W4-W5 regions in Cassiopeia \citep{lect78}.
Recently, \citet{wddv10} identified two different
age groups in the massive OB association Cyg OB2, while \citet{nb10}
could not find any signature of propagating star formation in the Vulpecula
OB association. On the other hand, \citet{pw10} suggested a possible
relation between star formation history and the IMF. They claimed evidence for
delayed massive star formation in the M17 SWex region from deviations in the
young stellar object (YSO) mass function from the Salpeter-Kroupa IMF.
Another issue relating to the IMF and star formation history is the IMF of
field stars such as the stars in the solar neighborhood.
Without detailed knowledge of the star formation history in a region
it is nearly impossible to trace the underlying IMF from the present day mass
function (see \citet{es06} for example).

This is the final paper based on extensive optical photometry of NGC 2264. 
The IMF around S Monocerotis and for low-mass stars in NGC 2264 was derived
in \citet{sbc04,sbc05} (Paper I \& II). \citet{sbc08} (Paper III) published
an extensive photometric catalogue of NGC 2264. They identified two active 
star forming regions (SFRs - S MON \& CONE) and a less active surrounding 
halo region (HALO).
More recently, \citet{ssb09} studied the spatial and spectral distributions
of young pre-main sequence stars in NGC 2264 based on {\it Spitzer} IRAC
(Infrared Array Camera - \citep{irac}) and MIPS (Multiband Imaging 
Photometer for {\it Spitzer} - \citep{mips}) 24$\mu m$ observations. From the spatial
distribution of Class I and II YSOs, \citet{ssb09} divided the CONE nebula
region of \citet{sbc08} into three SFRs - the Spokes cluster, Cone(C) 
and Cone(H).

This paper is organized as follows.
In \S2 we describe input data and some results from the SED fitting model 
(hereafter SED fitter).
The star formation history and the IMF of NGC 2264 are presented in \S 3 and
\S 4, respectively. We estimated the age and mass of stars in NGC 2264 in 
two ways, one from SED fitter, the other from color-magnitude diagrams (CMDs). 
In \S 5 we present a catalogue of young brown dwarf candidates (YBDCs) and 
their selection criteria. The effect of binarity on the IMF and
the IMF of M17 SWex by \citet{pw10} are discussed in \S 6.
The summary and conclusions are in \S 7.

\section{SPECTRAL ENERGY DISTRIBUTION MODEL FITTING}

In order to characterize the physical properties and the evolutionary status 
of disks and envelopes around pre-main sequence (PMS) stars in NGC 2264, 
we fitted the SED of stars detected in the mid-IR {\it Spitzer} observations
to the model grid of \citet{tpr06,tpr07} (we call this web page the SED 
fitter.)\footnote{\url{http://caravan.astro.wisc.edu/protostars/sedfitter.pho}}.
To obtain better results we compiled and used all the available optical
\citep{sbl97,psbk00,sbc08,ds05}, near-IR 2MASS \citep{2mass}, and mid-IR {\it 
Spitzer} data \citep{ssb09}. We limited our attention to objects which 
were detected in all IRAC bands and whose total photometric error in the 
four IRAC bands was smaller than 0.25 mag. The number of targets was 
1,173 (381 stars in S Mon, 185 stars in the Spokes cluster, 157 stars 
in Cone(C), 206 stars in Cone(H), 178 stars in Halo, and 66 stars in 
Field regions - see \citet{ssb09} for the region names).
Among them, 853 objects were known members (X-ray, H$\alpha$ emission, IR excess
from {\it Spitzer} IRAC, or early type stars) of NGC 2264. In addition, 33 
embedded objects in S Mon, Spokes cluster, Cone(C) and Cone(H) were also 
included. Among the above 886 objects, 336 objects (38 \%) were detected in
MIPS 24 $\mu m$ which gives an important constraint in SED fitting especially
for the YSOs with transition disks or with pre-transition disks.
The total number of known members in NGC 2264 is 1,162 stars, and therefore 
the fraction of stars used in SED fitting was 73.4 \% (= 853 / 1162) of 
the total known members. In addition we also examined physical properties 
of 40 BMS (below pre-main sequence) stars. 

SED fitter gives a wide range of models for a given SED. In such cases
we should have model selection criteria.
If several models give similar SEDs for a target, it is best to select
a model that gives the smallest mass for the central star because
(1) low-mass stars have a higher probability of having disks or envelopes,
and (2) low-mass stars occur more frequently than massive stars. 
In addition, in many cases, SED fitter gives a higher temperature than 
the temperature derived from the spectral type of the star (see \S 2.1.1).
In most cases the temperature of the $\chi^2_{min}$ model is equal to or
higher than that of the mass-minimum model. The best model 
selection criterion is therefore, the model which gives the smallest
mass for the central star among the 10 best models from SED fitter if $\chi^2 / 
\chi_{min}^2 \leq 2.0 $ or $\chi^2 < 2.0$ for the case of $\chi_{min}^2 
\leq 1.0$. But the best model for a BMS star is, on the other hand, the model 
whose $\chi^2$ is the minimum among the candidate models because if BMS stars 
are really PMS stars with edge-on disks, the criterion used for normal
stars above may give a biased result. 

In SED fitting we applied constraints on the distance and reddening, 
$\log d$ (kpc) = -0.15 -- 0.0 and $A_V$ = 0 -- 100 mag. The parameters
from SED fitter are those for the aperture of 2762 AU (about 3.$''$5 at 760 pc).

\subsection{Comparison of Stellar Parameters}

The physical or structural properties of PMS stars evolve as the central star
ages. We therefore expect an evolution or variation of such properties.
Young open clusters are ideal targets for this kind of study and
it is important to check these properties with the SED fitting tool. The
results can give some insights into an understanding of the structural evolution
of PMS stars as well as the evolution of the cluster itself. In addition,
it also provides evidence on the reliability of the SED fitting tool, 
and thereby contributes to improving SED fitter. 

In this section we compare several parameters relating to the central
star from SED fitter with those from other methods.

\subsubsection{Effective Temperature}

\begin{figure}
\epsscale{1.0}
\plotone{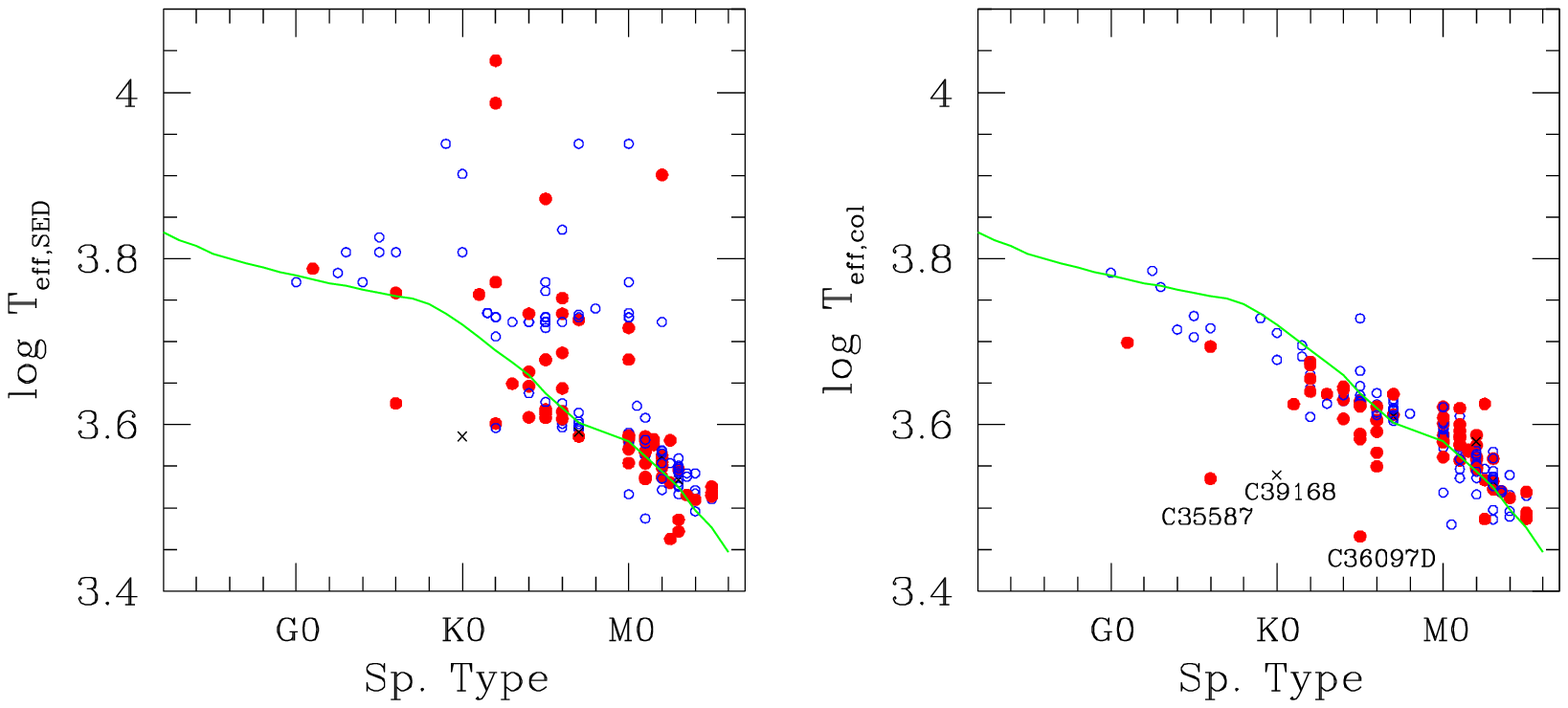}
\caption{Spectral type versus temperature relation. (Left panel) The effective 
temperature from SED fitting. (Right panel) The effective temperature is derived
from the $V-I$ or $R-I$ versus temperature relation of \citet{msb95}. 
We applied the same reddening for all stars ($E(B-V)$ = 0.07).
Spectral types are taken from \citet{lmr02}.
Dots, circles, and crosses represent, respectively,
Class II objects, stars with normal photosphere, and stars without
X-ray and H$\alpha$ emission. The solid line denotes the spectral type versus
temperature relation from \citet{lb82}. 
\label{sp_t} }
\end{figure}

Spectral type is the primary indicator of stellar temperature. \citet{lmr02}
published the spectral type of 360 stars in NGC 2264. Among them 174 stars are
targets for SED fitting. Figure \ref{sp_t} shows the relation between spectral
type and temperature. The left panel shows the relation between spectral type
and the temperature from SED fitting, while the right panel shows the same relation
but with temperature from the color versus temperature relation of \citet{msb95}.
As there is no known way to determine the reddening of late type PMS stars
using photometric data alone, we applied the same reddening $E(B-V)$ = 0.07
for all stars \citep{sbl97,psbk00}\footnote{We also applied a different 
reddening for a different SFR (the median value of reddening from SED fitter
in Table \ref{sedtab}), the result was very similar.}. The solid line in the figure 
is the spectral type versus temperature relation of MS stars \citep{lb82}.

Normally the expected error in spectral classification is 1 -- 2 subclasses, but
as many of our targets are Class II objects we can expect an error of 2 or more 
subclasses in spectral type due to the intrinsic variability. In the left panel of 
Figure \ref{sp_t} there are many stars with much higher temperatures from
SED fitter than that from the spectral type. Although we take the error in spectral
type into account, the difference in temperature is excessive. The reddening
of these stars would be more than 2.4 mag (mean $A_V =$ 3.32 mag). This value is
much larger than the median $A_V$ of all SFRs except Spokes cluster in Table 
\ref{sedtab}. This means that the reddening from SED fitter is less reliable,
and may cause a problem in the temperature estimate.

\placefigure{sp_t}

On the other hand the temperature derived from color indices is relatively 
well matched with that expected from its spectral type. There are three
exceptions - C35587, C39168, and C36097D (= C36097 + C36109). Two of them
are due to a large error in spectral type (G6: for C35587 and K0: for C39168).
C36097D may be due to duplicity or reddening correction (SED fitter estimates
the $A_V$ of C36097D as 6.26 mag).

The physical properties of the central source, such as effective temperature and
luminosity, are the most important parameters in understanding
the evolutionary status of the central source, and play a crucial role in 
the physical properties of disks and envelopes because dust in the disks and
envelopes are heated by irradiation from the star and from accretion 
shocks at the stellar surface. A large difference in temperature of some
stars in
Figure \ref{sp_t} implies that SED fitter seems not to use optical color as
a temperature indicator. As the optical color is a good temperature indicator,
it is better for SED fitter to estimate the temperature of the central source 
using optical color, such as $V-I$ which is least affected by UV excess
from accretion shocks or IR excess from dust in the disks.  In addition,
color is, in most cases, obtained from semi-simultaneous observations,
and therefore is less affected by the variability. As optical data, near-IR 
2MASS data, and {\it Spitzer} mid-IR data are obtained at different epochs, 
there are many cases showing a systematic difference in fluxes among
photometric bands due to variability. 

First determine the temperature of the central star from color indices, estimate 
the reddening $A_V$ from SED fitting, and then estimate the temperature again 
using reddening-corrected optical colors. This iterative process seems 
to be required to get better results.

\subsubsection{Mass and Age}

Unlike stars in the main sequence (MS) stage, the mass, as well as the age of PMS 
stars can be determined from the CMD with the aid of theoretical evolution models 
of PMS stars. As there is no reliable method of estimating the reddening 
for low-mass PMS stars without prior knowledge of spectral type, we have 
applied the mean reddening $E(B-V)$ = 0.07 mag
determined by \citet{sbl97,psbk00} using early type MS stars in NGC 2264.
This value is somewhat smaller than the median $A_V$ from SED fitter or
E($B-V$) from spectral type versus intrinsic color relation \citep{lmr02}.
But optically observable stars are relatively less embedded and
the difference in age or mass resulting from a smaller value of reddening
should be minor \citep{sbc04}. 

\begin{figure}
\epsscale{1.0}
\plotone{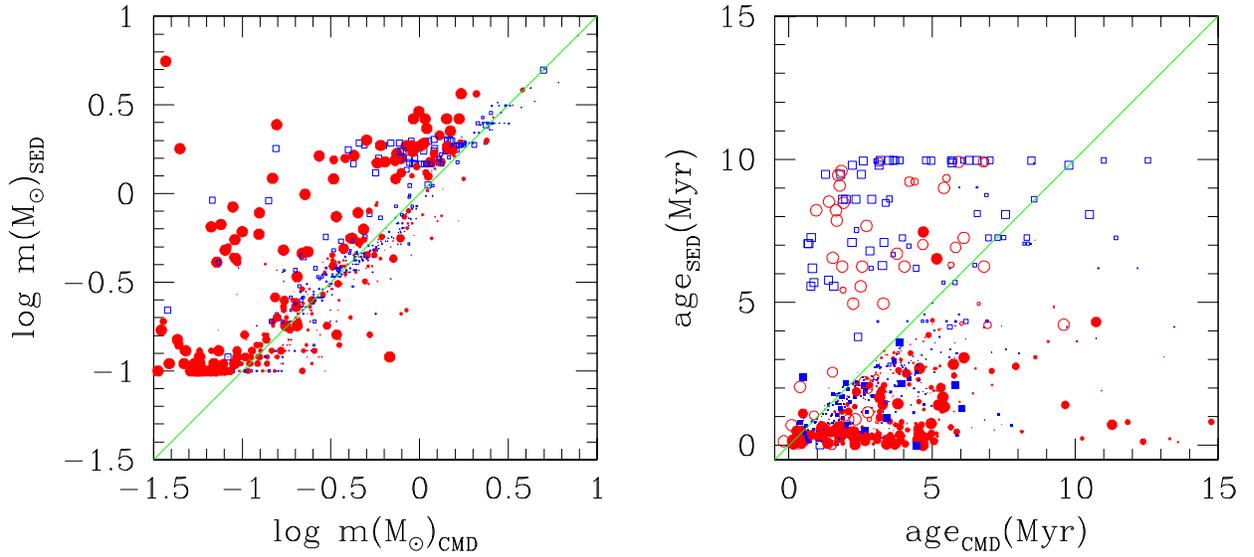}
\caption{Comparison of stellar mass and age. (Left panel) Comparison of stellar 
mass from
color-magnitude diagrams with that from SED fitter. The size of symbols is
proportional to the reddening from SED fitter. Dots and open squares represent
Class II objects and stars with normal photospheres, respectively. (Right panel)
Comparison of age from color-magnitude diagrams with that from SED fitter.
Circles and squares represent Class II objects and stars with normal 
photospheres, respectively. Filled and open symbols denote objects with 
M$_{\star,{\rm SED}} \leq$ 1 M$_\odot$ and  M$_{\star,{\rm SED}} >$ 1 M$_\odot$, respectively.
\label{mnt} }
\end{figure}

The PMS evolution models used are those by \citet{sdf00} for 
$m \geq 0.1$ M$_\odot$ and by \citet{bcah98} for $m < 0.1 $M$_\odot$ (see
Figure \ref{bdcmd} of this paper or Figure 4 of \citet{sbc04}). 
We compared the mass and age from CMDs with those from SED fitter in Figure 
\ref{mnt}. The size of symbols is proportional to the amount of reddening $A_V$ from SED fitter. 
Circles and squares represent Class II YSOs and stars with normal photosphere
(or Class III), respectively. In the right panel open symbols denote
massive stars ($m_{\star,{\rm SED}} >$ 1 M$_\odot$), while filled symbols 
stand for low-mass stars ($m_{\star,{\rm SED}} \leq$ 1 M$_\odot$).

\placefigure{mnt}

Masses of less reddened stars (small symbols) are well consistent with each other, 
but masses from SED fitter for high $A_{V,{\rm SED}}$ objects are in general
larger than those from CMDs. And we easily see the lower mass limit 
($m =$ 0.1 M$_\odot$) of SED fitter which is the mass limit of PMS
evolution models of \citet{sdf00}. In addition, there are two clumps
in the distribution of masses from SED fitter - a distinct clump at $\log m
\approx$ 0.2 -- 0.3 and a less prominent at $\log m \approx$ -0.2 -- -0.4. 
These two clumps
are directly related to the peaks in the IMF of Figure \ref{sedimf}.

The ages from SED fitter and from CMDs do not show any correlation (see 
the right panel of Figure \ref{mnt}). For low mass stars (m $\leq$ 1 
M$_\odot$), ages from SED fitter are in general smaller than those from CMDs, 
but for massive stars, ages from SED fitter are, on the other hand, larger 
than those from CMDs. The ages of massive stars from SED fitter are mostly 
larger than 5 Myr with an upper limit of 10 Myr, but those
from CMDs are distributed widely from $\sim$ 0 to 13 Myr. For low-mass stars
there is a strong concentration of ages from SED fitter between 0 -- 3 Myr.
But the ages from CMDs are distributed evenly between 0 -- 6 Myr. Many
Class II objects have relatively smaller ages from SED fitter. This fact
implies that the age difference may be partly related with the reddening correction.
But even when we applied the median value of reddening for a given SFR
the correlation does not improve noticeably. It is very
difficult to surmise the source of such differences.

\subsection{SED Fitting Results for Individual Stars}

As we are dealing with 
many PMS stars in the young open cluster NGC 2264 and expect that many 
parameters relating to the structure or evolution of PMS stars
are interrelated, it is therefore of interest to check 
whether there are any systematic variations of one parameter against another. 

\subsubsection{General Properties}

The mean value of the distance ($\log d$, $d$ in kpc) from SED fitting was -0.089 
$\pm$ 0.051 ($d$ = 815 $\pm$ 95 pc) for all PMS stars and the median value
was between -0.09 and -0.11 (see Table \ref{sedtab}). This value is
consistent with other determinations - 760 $\pm$ 90 pc from the zero age main 
sequence relation \citep{sbl97} and 913 $\pm$ 40 (sampling) $\pm$ 110
(systematic) pc from the assumption that the distribution of inclination 
angles determined from the projected rotational velocity and rotation period
is random \citep{ejb09}.

We could not find any correlation between age and reddening.
This means that the interstellar material in NGC 2264 may be very inhomogeneous.
The total luminosity did not show any systematic variation with age.
This is a result of a combination of the mass spectrum and different
evolutionary stages. If we limited our attention to the stars with masses
less than 1 M$_\odot$, we could find a weak evolutionary trend between
age and luminosity. The total luminosity of low-mass stars decreased
with age with the power of -0.28 $\pm$ 0.02. Other parameters did not show any 
systematic variation with age or mass.

\subsubsection{Disks}

\begin{figure}
\epsscale{0.8}
\plotone{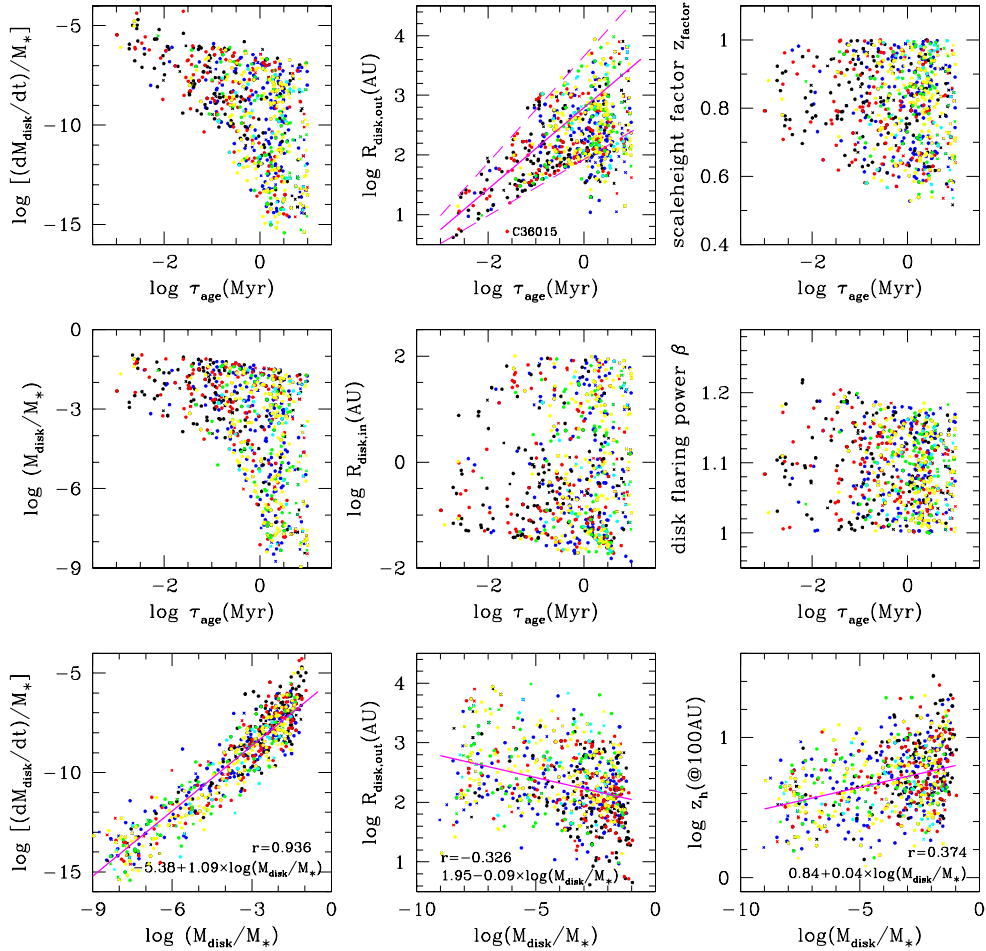}
\caption{Disk parameters of individual stars from SED fitter. 
The upper two panels show the time evolution of several parameters relating 
to disks around PMS stars, while the lower panel displays the variation of 
several parameters against disk mass. To see whether there are any 
systematic differences among SFRs or evolutionary stages we use different 
colors and different symbols. Dots and crosses represent, 
respectively, Class I or II objects \citep{ssb09} and stellar photospheres.
Blue, black, red, green, yellow, and cyan represents the PMS stars in S Mon, 
Spokes cluster, Cone(C), Cone(H), Halo, and Field, respectively. \label{seddisk} }
\end{figure}

\placefigure{seddisk}

We present several figures relating to the disk properties in Figure 
\ref{seddisk}. We show the relation between stellar age and disk accretion
rate in the first panel of Figure \ref{seddisk}. There seems to be virtually
no variation in the upper limit of the disk accretion rate, but 
the lower limit decreases with time. The scatter in the disk accretion 
rate of $\log \tau_{age} \lesssim -1$ is about 3 orders of magnitude, 
rising to about 8 orders of magnitude at $\log \tau_{age} \approx +1$.
The relation between disk mass and stellar age shows a similar variation 
with time and a similar scatter. As a result, we see a rather tight 
correlation between disk mass and disk accretion rate.

$$ \log (\dot{M_{disk}}/M_\star) = -5.38 (\pm 0.06) + 1.09 (\pm 0.01) \cdot 
\log (M_{disk}/M_\star ), ~(r = 0.936). $$

\noindent
There is an order of magnitude scatter around the relation at a given disk mass.
In addition, the relation shows a slight curvature - higher accretion rates
for massive disks. Stars in the younger SFRs - Spokes cluster and Cone(C) - 
are more concentrated in the upper right of the figure, i.e. higher accretion 
rates and more massive disks.

The $R_{disk,out}$ of individual stars increases with age. The power of the lower
boundary is about 0.47 and that of upper boundary is about 0.89. The initial 
scatter ranges over about an order of magnitude, but it increases with age 
and reaches about two orders of magnitude near $\log \tau_{age} \approx 1$. 
And there are several outliers between $\log \tau_{age} =$ 0 -- 1 as in
Figure 6 of \citet{tpr06}.  Some of these outliers are double stars, which is
reminiscent of disk truncation due to the existence of companion, but it is
uncertain why they have smaller $R_{disk,out}$ than the other YSOs.
There is an exception which cannot be found in Figure 6 of \citet{tpr06}.
 C36015 has very small $R_{disk,out}$. C36015 is a Class I object
close to the brightest IR source in NGC 2264 IRS 1 ($0.'362$ = 0.08 pc at 
760 pc).

If disk evolution occurred in a gradual 
manner, we could expect an increase in the disk inner radius with age.
But the inner disk radius ($R_{disk,in}$ in units of $R_{sub}$ 
or AU) for individual stars shows no systematic variation with age.
The disk scale height at 100AU, on the other hand, is greater for younger
stars with massive disks ($\log z_{h,100AU} = 0.84 (\pm 0.02) + 0.04 
(\pm 0.00) \cdot \log (M_{disk}/M_\star ), ~r = 0.374$). 
Other parameters, such as scale height factor z$_{\rm factor}$,
flaring factor $\beta$ and disk accretion power $\alpha_{acc}$ 
show no appreciable systematic variation with time. There is a very small
variation of z$_{\rm factor}$ and  $\beta$ with time, and \citet{tpr06}
interpreted such variations as the settling of dust to the disk mid-plane.

\subsubsection{Envelopes}

\begin{figure}
\epsscale{0.8}
\plotone{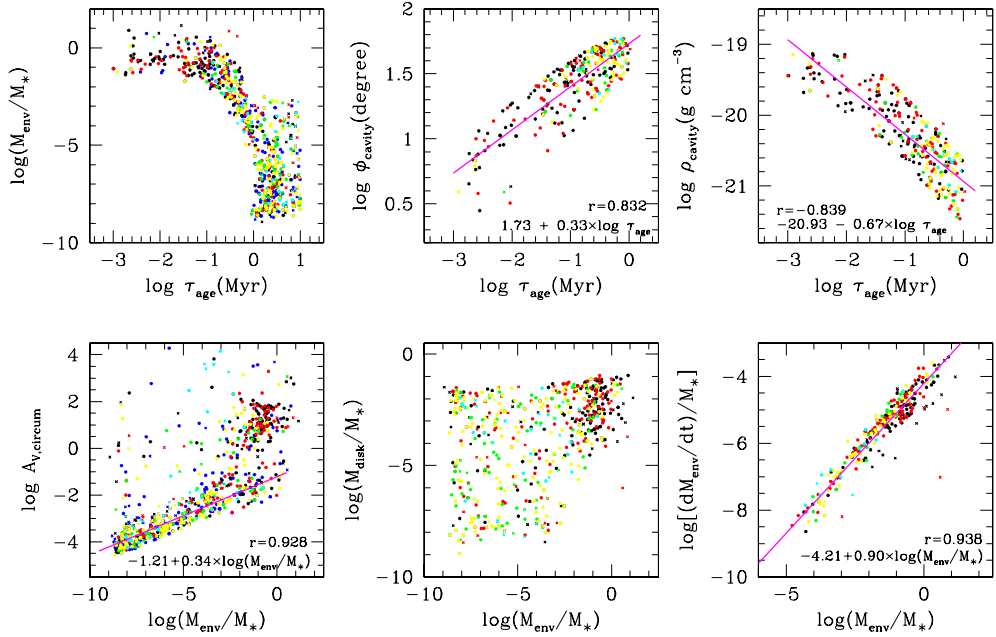}
\caption{Envelope parameters of individual stars from SED fitting. 
The upper panel shows the time evolution of several parameters 
relating to the envelopes. The lower panel shows the relation between 
circumstellar reddening, disk mass, or envelope accretion rate and
the mass in the envelope. Symbols are the same as in Figure \ref{seddisk}.
\label{sedenv} }
\end{figure}

\placefigure{sedenv}

We present several figures relating to the envelope parameters in Figure 
\ref{sedenv}. The mass in the envelope is about 10$^{-1 \pm 1}$ M$_\odot$
at the initial stage and decreases gradually with time reaching about 10$^{-4
\pm 1}$ M$_\odot$ at about 1 Myr. After that, the envelope masses show a large
scatter. The mass in the envelope disperses rapidly and accretion 
from the envelope is halted at this stage (see the last figure of Figure
\ref{sedenv}). However, the large scatter in the envelope mass 
at $\tau_{age} \gtrsim$ 1Myr could simply be an artifact of the SED fitting.

Although there is a large scatter, older stars have a larger cavity opening 
angle ($\log \phi_{cavity} \approx 1.73 (\pm 0.01) + 0.33 (\pm 0.01) \cdot 
\log \tau_{age}, ~r = 0.832$). The cavity opening angle is 
about 10$^\circ$ at 0.01 Myr, and increases up to about 55$^\circ$ at 1 Myr. 
The cavity density, on the other hand, decreases as age increases
($\log \rho_{cavity} \approx -20.93 (\pm 0.02) - 0.67 (\pm 0.02) 
\log \tau_{age}, ~r=-0.839$). The initial value of the cavity density is
about 10$^{-19.5 \pm 0.5}$ g cm$^{-3}$, and decreases with age to the power
of -2/3, i.e. $\rho_{\rm cavity} \propto \tau_{\rm age}^{-2/3}$.
At an age of 1 Myr, the cavity density approaches about 10$^{-21.0 \pm 0.5}$
g cm$^{-3}$. This fact implies that the sweeping up of envelope material
occurs at a very early stage.

The circumstellar extinction $A_{V,circum}$ (the extinction from the edge of the
circumstellar envelope to the surface of the central star) decreases 
with stellar age as $A_{V,circum}$ $\propto \tau_{\rm age}^{-1.75 \pm 0.06}$.
There are three distinct groups in the diagram of envelope mass and
circumstellar extinction. The first is the highly extincted
objects with massive envelopes, the second group is less extincted with less massive
envelopes. The second group show a good correlation between envelope mass
and circumstellar extinction ($\log A_{V,circum} \approx -1.21 (\pm 0.03) + 
0.34 (\pm 0.01) \log (M_{env}/M_\star ), ~r=0.928$). The first group comprises
PMS stars with active accretion from the envelope. The third group comprises relatively highly
extincted objects regardless of envelope mass. These are
BMS stars with nearly edge-on disks (see Table \ref{sedtab}).

It is natural to expect a correlation between envelope mass and disk mass, 
but the results show no such correlation between these quantities. There
is a strong concentration at $(M_{env}/M_\star ) \approx 10^{-1.0 \pm 1.0}$
and $(M_{disk}/M_\star ) \approx 10^{-2.5 \pm 1.5}$ and there is a 
scattered population with $(M_{env}/M_\star) \lesssim 10^{-4}$. 
The stars in the concentrated group are mostly stars in Spokes cluster
or Cone(C). Their envelope mass of the scatter population may be 
an artifact of SED fitting as mentioned before.

The envelope accretion rate is a strong function of envelope mass
($\log (\dot{M_{\rm env}}/M_\star ) = -4.21 (\pm 0.04) + 0.90 (\pm 0.02) \cdot \log
(M_{\rm env}/M_\star ), ~r = 0.944$). Many active stars in the Spokes 
cluster or the Cone(C) region 
show an envelope accretion rate of $10^{-5.8 \pm 0.5}$ M$_\odot ~yr^{-1}$.
The time evolution of the envelope accretion rate obtained here is very similar 
to that in Figure 4 of \citet{tpr06}.
The outer radius of the envelopes does not show any correlation with age or mass.
And there are only a handful of stars with $R_{env,out} \geq$ 1000 AU.

\subsection{Median Properties}

\placetable{sedtab}

We present the median values from SED fitter in Table \ref{sedtab}. 
Although the membership selection is far from complete in the Field region, we list 
the data for completeness. As the scatter of a given parameter is very 
large due partly to the age spread in a given SFR and partly to the 
difference in environments, the values in the table are (logarithmic value of)
median, 10 and 90 percentiles except the first two rows which give the fraction
of objects in a given evolution stage.  As shown in the 5th row of Table 
\ref{sedtab} the median age changes from 0.35 Myr to 2.02 Myr. But if we define
the age spread of a given SFR as the age difference between the 10 
and 90 percentiles in the cumulative age distribution \citep{psbk00}, the age
spread of SFRs in NGC 2264 is between 2.5 Myr and 3.4 Myr, which is 
larger than the difference in median ages among SFRs. This means that
most properties of YSOs in a given SFR in parameter space overlap with those
in other SFRs. Although we have such a limitation, we adopt the median age
as the representative age of a given SFR. In this section we examine
the variation of parameters among SFRs against their median ages. This picture
can give an approximate view of several young open clusters with different
ages.
 
\subsubsection{General Properties}

The first two rows of Table \ref{sedtab} list the fraction of objects
at a given evolutionary stage.
The definitions used in \citet{tpr06} for the evolutionary stages are adopted
and used here. The fraction of stars in Stages 0/1 or 2
is much higher than that of stars in Class I or II stages in \citet{ssb09}.
Such a difference is due mostly to the fact that only members or probable
members (i.e. embedded objects in the three active SFRs) are taken into account.
The difference in the definition of ``class'' and ``stage'' of YSOs also 
contributes to the difference. As already mentioned in \citet{ssb09}, the 
evolutionary stage of SFRs in NGC 2264 differs from one region to another, 
so we can expect difference in the age of each SFR.
The age in Table \ref{sedtab} is the median, 10 and 90 percentile age of stars 
whose mass is between 0.2 -- 1.0 M$_\odot$ because most PMS evolutionary models
give an abnormally large age for massive PMS stars \citep{sbl97,scb00,sbc04} (see
also \citet{lh03}). The Spokes cluster is the youngest ($\tau_{median} = 0.35$ Myr),
S Mon, Cone(H), and Halo are slightly older ($\tau_{median} =$ 1.6 -- 2.0
Myr), and Cone(C) is intermediate between them ($\tau_{median} = 0.76$ Myr).
More discussions on the age of stars and star formation history in NGC 2264
are made in \S 3.

\placefigure{sed_median}

To see whether there are any systematic variation in physical parameters
among SFRs, we plot several parameters against the median age of a given SFR
in Figure \ref{sed_median}. The median value of interstellar extinction
against median age is shown in Figure \ref{sed_median} (1). 
Although we cannot find a noticeable trend for individual stars in a given 
SFR, the median relation shows a clear evolutionary
trend between extinction A$_V$ and age. Despite the small samples,
young stars in the Field region or BMS stars deviate largely from the trend. 
Field members are far away from the active SFRs in NGC 2264 and so are more 
likely to be in relatively transparent regions. The median A$_V$ of BMS stars are
slightly larger than the others. It is reasonable that BMS stars have a larger
value of A$_{V,~circum}$ as they are thought to have
nearly edge-on disks. But the larger A$_V$ of BMS stars may be just a result
of small sample statistics or result from difficulties in fitting the SED of 
BMS stars.

The median value of the total luminosity shows a strong correlation with age.
As there is no correlation between age and mass, the correlation between
age and total luminosity reflects the average evolutionary stage of a given 
SFR. A similar
figure for individual stars in a given SFR shows a similar trend with
a large scatter. Such a scatter is due to the diversity of stellar masses
as well as to a difference in evolutionary stage. BMS stars show higher 
luminosity at a given age. That is probably due to the higher dust emission
from disks around BMS stars as the median value of the disk masses of BMS stars
is higher than for other stars. 

\subsubsection{Disks}

Several of the next figures show the behavior of parameters relating to disk
properties. The median value of the inclination angle of disks is the same
for all SFRs and the distribution of inclination angle of individual stars
for a given SFR does not show any preferential angle, i.e. a random distribution 
of inclination angles. On the contrary, the median value of the inclination angle 
of BMS stars is very high (the highest inclination angle from the SED fitter). 
This result is partly due
to the difference in the best model selection criterion. If we use the same
criterion for the best model for BMS stars, the median value of the inclination
angle is 69.5$^\circ$ $_{40.5}^{87.1}$. A Kolmogorov-Smirnov (K-S) test between
the cumulative distribution of inclination angles of PMS stars in the S Mon 
region and that of BMS stars gives 6.5 $\times 10^{-2}$ \% probability of 
a subset drawn from the same underlying distribution. This value is far lower 
than the probability between the S Mon region and the Spokes cluster (89.7 \%) 
or the S Mon and Cone(C) regions (29.7 \%). This result also implies that 
BMS stars are PMS stars with nearly edge-on disks.

The mass in the disks as well as the mass accretion rate are strong functions of age,
i.e. PMS stars in younger SFRs have larger disk masses and higher disk accretion 
rates as expected.  
The median value of the disk inner or outer radii is, on the other hand, smaller
for younger SFRs and larger for older SFRs. The $R_{disk,out}$ of individual 
stars in an SFR show a similar trend as shown in Figure \ref{seddisk}, 
but there is about two orders of magnitude difference between minimum and 
maximum value of $R_{disk,out}$ at a given age. The inner radius ($R_{disk,in}$
in unit of $R_{sub}$ or AU) of individual stars in an SFR does not show 
any trend as mentioned in \S 2.2.2. 
Other parameters, such as scale height factor z$_{\rm factor}$, 
flaring factor $\beta$, and disk accretion power $\alpha_{acc}$ 
do not show any appreciable systematic difference or variation among SFRs.
But the disk scale height at 100AU decreases as age increases.

We also show the relation between disk mass and other parameters in
Figure \ref{sed_median} (21) -- (25). The disk accretion rate is a strong
function of disk mass. The relation between the median value of the disk masses
and the median value of the disk accretion rate is

$$ \log \dot{M_{disk}^{med}} = -4.76 (\pm 0.35) + 1.28 (\pm 0.09) \log M_{disk}^{med},  ~~(r = 0.990) $$

\noindent
Disk mass and inner or outer radii are anti-correlated in the sense of median
values. The intercept and exponent powers differ from the same 
relations for individual
stars. The median value of the total luminosity is proportional to the median 
value of disk mass even for BMS stars, as seen in Figure \ref{sed_median} (24). 
But due to the variety of disk masses and evolutionary
stages there is no correlation between disk mass and total luminosity for
individual stars. The median value of the disk scale height at 100 AU is well
correlated with the median value of the disk mass (Figure \ref{sed_median} (25)).
This result means that although there is nearly no change in scale height 
factor $z_{factor}$, the disk scale height at 100 AU is larger for massive 
disks, in other words it is larger at earlier stages.

\subsubsection{Envelopes}

The next figures (Figure \ref{sed_median} (13) -- (18), (26) -- (30)) show 
the time evolution of parameters relating to the envelopes of PMS stars 
in NGC 2264. It should be noted that in most cases the envelope could be
observed for those stars with age $\lesssim$ 1 Myr, therefore the median
age of a given SFR cannot represent the age of individual stars with envelopes.

The cavity opening angle increases as the median age increases.
The decrease in median value of envelope mass with age is more dramatic being 
on average, a four orders of magnitude decrease in envelope mass from the youngest
SFR Spokes cluster to the $\approx$ 2 Myr SFRs S Mon, Cone(H) and Halo regions.
The difference in envelope accretion rate among SFRs is small - only an approximate 
5 times difference between the Spokes cluster and the S Mon or Halo regions. 
The envelope accretion
rate with envelope mass of individual stars changes in a similar manner.
The exponent of the power in the median relation (Figure 
\ref{sed_median}(26)) is 0.14, but that of individual stars is 0.85

The median value of the envelope outer radii is well correlated 
with the median age of the SFR as shown in Figure \ref{sed_median} (14).
This may be due to the infall of material to the disk.
For individual stars there is almost no correlation between the envelope outer 
radius and stellar age as mentioned in \S 2.2.3. 
The median value of the inner radius of the envelope increases as the median age 
increases as shown in Figure \ref{sed_median} (15). 
But any correlation between them is not so clear, 
because SED fitter gives the smaller, default value of the
inner radius of the envelope ($R_{env,in}$ = 1 $R_{sub}$) for many cases.

The median value of the density in the cavity decreases with median age 
of the SFRs. But the median value of the ambient density $\rho_{\rm ambient}$ 
is nearly constant regardless of the age of the SFRs.
BMS stars in Figure \ref{sed_median}(19) show a larger $\rho_{\rm ambient}$.
As there is no reason to expect higher $\rho_{\rm ambient}$ for BMS stars,
this may be the result of difficulties of SED fitting for BMS stars.
The circumstellar extinction (A$_V$ from the outside edge of the YSO to the
stellar surface along the line of sight) is a decreasing function of age
as is the interstellar extinction in Figure \ref{sed_median}(1).
The median value of the circumstellar reddening A$_{V,circum}$ 
decreases as the median age increases with about the -4th power of the age.
The A$_{V,circum}$ of BMS stars is very high due to extinction
in the circumstellar disk, and does not appear in Figure \ref{sed_median}(20).

Although there are strong correlations between the median value of envelope mass
and the median value of disk mass or total luminosity as shown in Figures \ref{sed_median} (29)
and (30), there is no correlation between the same quantities for individual stars.

\section{STAR FORMATION HISTORY}

The star formation history in an SFR can give valuable information on star
formation processes. From the age difference between massive stars near the
MS turn-off and low-mass PMS stars near the MS turn-on, \citet{gh62} proposed
a sequential star formation scenario in an open cluster. But \citet{sbl97}
claimed that the mass-age relation of PMS stars in young open clusters
may be an artifact of PMS evolution models, and suggested that such 
a discrepancy in age could be solved by reducing the PMS lifetime (Helmholtz
-Kelvin contraction time scale). Later 
\citet{scb00,psbk00,sbc04} supplied additional supporting results. However 
\citet{ps00} proposed prolonged star formation which is nearly the same
scenario as suggested by \citet{gh62}. 

\citet{gf06} identified three groups in NGC 2264 according to their radial 
velocity from high resolution multi-object spectroscopic observation. 
Later \citet{sbc08} confirmed the existence of two of them and found one 
more SFR, Halo, surrounding these SFRs. \citet{pst06} first identified 
the Spokes cluster as being dominated by Class I sources. Using {\it Spitzer} 
mid-IR data \citet{ssb09} also identified two SFRs (Spokes cluster and Cone(C))
dominated by Class I objects and one SFR (S Mon) dominated by Class II 
objects. Throughout this paper we use the name of the SFRs used by 
\citet{ssb09}. It is very interesting to now compare the age distribution 
among SFRs. It should also give some important information relating to 
the star formation processes within NGC 2264.

\subsection{Age Distribution}

The age of PMS stars is determined in two ways. One is the age from the
SED fitter, and the other is the age from the CMDs (see \S 2.1.2).
We present the cumulative age distribution of 
low-mass PMS stars in a given SFR in Figure \ref{figsfh}.
To avoid unrealistic ages from PMS evolution models, we limit the PMS
stars with m = 1.0 -- 0.2 M$_\odot$ (see \citet{sbl97,sbc04}). 

We also present the median age, 10 and 90 percentile in the cumulative
age distribution, and the results from K-S tests
in Table \ref{agetab}. If we assume that the median age is the representative
age of an SFR and the difference between the 10 and 90 percentile in the age
distribution is the age spread of the SFR, the median age from SED fitter
is about 1 -- 2 Myr younger than that from CMDs for a given SFR.
In addition, the age spread is about 3 Myr from SED fitter, while
that from CMDs is about 5 -- 6 Myr. The smaller value of age spread
is due to the fact that the stars used in SED fitter are relatively
younger and easier to detect with {\it Spitzer} observations.
On the other hand, the somewhat larger age spread from CMDs may be due to
the effect of reddening. In addition,
the reddening law in young SFRs may differ from that in the general
interstellar medium. Therefore, the age spread, or in other words the  
cluster formation time scale, may be about 5 Myr.

\placefigure{figsfh}
\placetable{agetab}

The left panel of Figure \ref{figsfh} shows the cumulative age distribution
of PMS stars from SED fitter. Although the median age of stars in Field
is much younger than that in Halo, this seems to be due to the small number of
member stars in Field. As the age distribution of stars in Field is very
similar to that of Halo and the number of known members with H$\alpha$ emission
or X-ray emission in Field is very small, we merge the stars in Field and Halo.
The age distribution of stars in a given SFR is far different from each other. 
The stars in the Spokes cluster are the youngest among the SFRs in NGC 2264. The stars 
in S Mon and Cone(H) have a similar distribution, while the age of stars in 
Cone(C) is intermediate between the Spokes cluster and S Mon. The age of stars 
in Halo is slightly older than that in S Mon, but the difference is marginal.
To check whether the age distribution among SFRs in NGC 2264 arise
from the same distribution or not, we have performed a K-S
test. The results are listed in Table \ref{agetab}. The probability
of the subsets being drawn from a common population is very low between
S Mon and the Spokes cluster or S Mon and Cone(C), but is very high between S Mon and 
Cone(H). The probability between S Mon and Halo (and Field) is marginal.

The cumulative distribution of ages from the CMDs is presented in the right panel
of Figure \ref{figsfh}. The pattern of age distributions is far 
different to that from SED fitter. The age distributions of all active SFRs 
(S Mon, Spokes cluster, Cone(C), and Cone(H)) in NGC 2264 are very similar.
As many young PMS stars in the Spokes cluster or Cone(C) are deeply embedded, they
are not in the PMS locus of NGC 2264. In addition, the PMS stars in the PMS
locus are situated in the outside surface region of NGC 2264, and therefore
the contamination of Halo or Field members is unavoidable. This is the reason
why the age distribution of these active SFRs is similar to each other.
On the other hand the difference in age distribution between stars in S Mon and
Halo (including Field) is rather clear. The probability from a common 
population is very low ($P = 4 \times 10^{-4}$). Somewhat larger probabilities 
relative to the age distribution of stars in the Spokes cluster are obtained 
from the K-S test. The results are due to the small number of member stars 
in the PMS locus of NGC 2264 because most of the members of the Spokes cluster 
are an embedded population. In order to check the reliability of this 
comparison, although there is no way to estimate the age of non-member 
stars, we estimated the age of 
non-member stars in the PMS locus of NGC 2264 by assuming they were members 
of NGC 2264. The age distribution of non-member stars is far different 
from that of Halo or other member stars in active SFRs. The very low 
probabilities from a K-S test between the age distribution of non-member 
stars in Field and other SFRs indicate that this kind of test is 
valid for the study of star formation history.

\subsection{Triggered Star Formation in NGC 2264}

The age distributions obtained above give a clear 
picture of sequential star formation
in NGC 2264. The stars in Halo (and Field) region are formed first, 
followed by stars in S Mon and Cone(H). The stars in Cone(C) 
formed later and finally star formation took place
in the Spokes cluster. 
We can find several Class I objects in Cone(C), and many Class I objects
\citep{ssb09} and Class 0 objects \citep{tzl07} in the Spokes cluster. In other words, 
star formation activity in NGC 2264 started at the surface of the giant Monocerotis 
molecular cloud and propagated inside the cloud. As already noted in 
\citet{sbc08} there are several small SFRs spread over Field region. 
The median age of Field is very similar to that of Halo. It is very difficult
to imagine that spontaneous star formation started first in a low
density region (Halo or Field) of NGC 2264. If the star formation history
derived in \S 3.1 is true, we should accept that 
star formation in NGC 2264 was triggered by an external source, such as 
a supernova (SN) explosion that occurred about 6 -- 7 Myr ago.

An extensive review of triggered star formation has been published 
by \citet{elm98}. He classified triggered star formation 
into three types according to the scale of star formation. \citet{tlg93} 
conducted a high resolution CO observation in the bright nebular region 
of NGC 2264 and found three bright clumps. This is an example of small 
scale triggered star formation in NGC 2264. The age difference between
S Mon and Cone(C) is 0.4 -- 0.8 Myr, and the projected distance between
S Monocerotis and HD 47887 is about 26$'$ (= 5.7 pc at 760 pc). 
If star formation in Cone(C) were triggered by the influence of
S Monocerotis, the minimum propagation velocity of star formation activity
would have been 7 -- 14 km s$^{-1}$, which is far faster than sound speed
in a general molecular cloud ($\sim$ 1 km s$^{-1}$). Therefore intermediate 
scale triggering - ``collect and collapse'' - may not have taken place in 
NGC 2264.

The first mention of large scale triggered star formation happening in 
and around NGC 2264 is the paper by \citet{ch79} who found an HI shell
in the Monocerotis region. Later \citet{prs87} found a half circle stellar loop
in Mon OB1 region using IRAS images. He listed several young objects
along the semicircle - IC 446, IC 2169, NGC 2264 IRS1 and IRS2, and NGC 2261 
(= R Mon). The diameter of the half circle was about 2.5 degree (about 33 pc
at 760 pc). \citet{omt96} studied the large scale structure in Mon OB1
from an extensive CO survey. They identified many molecular clouds and listed
candidate optical objects possibly associated with them. The center of the
shell is ($l$, $b$) = ( $\sim 202^\circ$, $\sim 2^\circ$) from Figure 1
and 2b of \citet{omt96}.

\subsection{Objects of Common Origin with NGC 2264}

We showed in the previous section that the star formation in NGC 2264 was 
triggered externally, and the triggering source may have been an SN that 
exploded about 6 -- 7 Myr ago.
\citet{prs87,omt96} listed many possible optical counterparts.
We have checked these objects and a few should be excluded.
All data were obtained from the Simbad database \footnote{
\url{http://simbad.u-strasbg.fr}} and the open cluster database WEBDA 
\footnote{\url{http://www.univie.ac.at/webda}}.
There are no optical counterparts for approaching clouds 4, 12, and 19 in 
Figure 2a of \citet{omt96}. Although the large extent in velocity of cloud 4 
in Figure 1 of \citet{omt96} implies the possibility of it being an ejected cloud 
from the imputed SN explosion, the cloud is located far from the center of the
CO hole ($l \approx 202^\circ$, $b \approx 2^\circ$). Cloud 4 may therefore
not have the common origin. Cloud 19 may also not be related to the SN
explosion for the same reason. Cloud 9 shows a large velocity gradient and
is very close to cloud 10. There are two reflection nebulae (IC 2167 = IC 446;
GN 06.28.4) in cloud 9, and several young variable stars in and around these
 reflection nebulae. These young stars are 2MASS J06310611+1027340 (Sp: B2.5V) 
in GN 06.28.4, VY Mon (Sp: B8, H$\alpha$ emission star) in IC 2167, V540 Mon
(= HBC 521, Sp: M6, H$\alpha$ emission star) and V687 Mon (= LkH$\alpha$
274, Sp: K4IVe) around IC 2167.

Cloud 10 is very close to the CO hole and is an approaching cloud. 
Within it are an HII region IC 2169 and 4 early type stars - HIP 31038
(= vdB 76, Sp: B7IIIp), V727 Mon (Sp: B8V, eclipsing binary), HD 258749 
(= vdB 77, Sp: B5III), and HD 258853 (= vdB 78, Sp: B3).
There are two reflection nebulae in cloud 11 - NGC 2245 and NGC 2247. 
An H$\alpha$ emission star V699 Mon (= LkH$\alpha$ 215, Sp: B6) is at 
the center of NGC 2245, and a B6pec star V700 Mon is in NGC 2247.
Although cloud 11 is a very massive cloud, only two reflection nebulae
indicate current star formation activity. This may be due to its small
cross section facing the CO hole.

Cloud 13 is related to IRAS 06337+1051, many radio sources and a thin 
nebula around the mid-IR source. Several young or massive star 
candidates can also be found. Among them Kiso H$\alpha$ 39 may be related to
the cloud, but the relation of GSC 00737-01170 (Sp: O5:) and Tycho 
737-898-1\footnote{$V$ = 10.18, $B-V$ = 1.22, $U-B$ = 0.42 by 
\citet{jsd75}. The $B-V$ color is too red for a B type star. If the star 
were a low-mass PMS star and its blue color caused by UV excess, the star 
would be somewhat brighter than the low-mass members of NGC 2264.} 
(Sp: B) to cloud 13 is uncertain. \citet{prs87,omt96} proposed
a possible relation between cloud 13 and the open cluster NGC 2259.
However, NGC 2259 is an intermediate-age open cluster that cannot 
be related to the cloud.
The most massive molecular cloud in Table 1 of \citet{omt96} is the giant
molecular cloud
which is associated with NGC 2264. A small cloud - cloud 17 - harbors 
a reflection nebula NGC 2261 and the young star R Mon (Sp: B0).

\citet{omt96} proposed a possible relationship between the open cluster
NGC 2254 and clouds 18 through 20. NGC 2254, as well as Basel 7 and Basel 8, have
been observed only in the photographic $RGU$ system  and so their nature is 
somewhat uncertain. They appear to be intermediate-age open clusters from 
their CMDs. Another open cluster NGC 2251 is marked in Figure 2b of 
\citet{omt96}. The cluster's distance  in WEBDA is more distant than 
NGC 2264, and although the data were derived mostly from $UBV$ photographic 
photometry the cluster seems to be a Pleiades-age open cluster.
Therefore the relationship between the clusters and the clouds is quite 
uncertain. New homogeneous data are required for a comprehensive view 
of the star formation history in the Mon OB1 association.

\section{THE INITIAL MASS FUNCTION}

In this section we derive the IMF of NGC 2264 in two ways - using masses derived
from SED fitter and masses derived from fitting CMDs.  
The membership selection is nearly complete down to 
about 0.25 M$_\odot$ \citep{sbc04} for two {\it Chandra} fields - one around
S Monocerotis \citep{svr04} and the other around the Cone nebula \citep{fl06}.
These two fields fully cover the four active SFRs (S Mon, Cone(C), Cone(H), and 
Spokes cluster) in NGC 2264. We designated these four regions as the reference 
region for the combined IMF. The IMF is defined as the number of stars per unit 
projected area per unit logarithmic mass interval, i.e. $\xi (\log m) = 
d N / \Delta \log m / $area, where area is in kpc$^2$. We adopted the distance of
NGC 2264 as 760 pc \citep{sbl97}. We used the logarithmic mass interval 
$\Delta \log m$ = 0.1 for the reference region, and $\Delta \log m$ = 0.2
for the IMF of an individual SFR. We also calculated the IMF for each 
SFR in the same bin size, but
shifted by 0.1 in $\log m$ to reduce the binning effect. For massive stars 
($\log m \geq 0.5$) there were many cases with no star in a given mass bin. 
In those cases, we enlarged the mass interval so that there was at least one 
star in
each mass bin. The upper mass limit was assumed to be 100 M$_\odot$
 - the most massive star ever measured was 116 $\pm$ 31 M$_\odot
$ \citep{scc08} - and we calculated the mean value of $\log m$ 
in the mass bins. One thing to be noted is that although some stars were
known to be in a binary system, such as S Mon \citep{drg97,sbl97}, we assumed
all stars to be single as we have only limited information on multiplicity
(see \S 6.1 for the effect of binarity).

\subsection{The IMF from SED Fitter}

We determined the IMF for each SFR, but show the IMF of 
the reference region in the left panel of Figure \ref{sedimf} because a) 
membership selection is nearly complete; b) we could determine the slope of 
the IMF in the massive regime (M $\geq $ 3 M$_\odot$) reliably as only a few 
massive stars were in a given SFR. The right panel of Figure \ref{sedimf} shows
the relative IMF of each region relative to the reference IMF.
The abrupt decrease in the IMF at $\log m \leq$ -1.0 is due to the limited
exposure time of the {\it Spitzer} observations of NGC 2264. There are some 
stars with $\log m <$ -1.0 although the lowest mass from SED fitter is 
0.1 M$_\odot$;  the mass of these stars is derived from CMDs only.

\placefigure{sedimf}

The IMFs by \citet{kro02} and by \citet{mbsc03} are superimposed for 
comparison. Their IMF were adjusted roughly at $\log m =$ -0.3 -- -0.4.
There is a distinct peak in the combined IMF of NGC 2264 at about 2 
M$_\odot$ and another less distinct peak at about 0.4 M$_\odot$ as already
mentioned in \S 2.1.2. We could not find similar peaks in the IMF of field 
stars in the Solar neighborhood \citep{jms86,ls95,cha01,kro02} or in young 
open clusters \citep{sbc04,mlla02}.
In many cases there was a local maximum at about 1 M$_\odot$ in the IMF 
of field stars or open clusters, therefore the prominent peak at
M $\approx 2M_\odot$ in Figure \ref{sedimf} may be an artifact induced by 
the $\chi^2$ minimizing routine in 
SED fitter. Such a peak can be seen the YSO mass function of M17 SWex (see
\S 6.2 for the discussion on this issue) \citep{pw10}. 
Normally the IMF can be characterized by its slope $\Gamma (\equiv {d \log \xi}
/ {d \log m})$. The slope for massive stars ($\log m \geq 0.5$) is
-1.79 $\pm$ 0.14, which is very similar to the slope obtained by \citet{psbk00}
for NGC 2264 or by \citet{jms86} for the Solar neighborhood.

Figure \ref{sedimf} (right panel) shows the difference in the IMF for each
SFR relative to the combined IMF in the left panel. The IMF itself, 
by definition, represents the surface density. The surface density
of stars in the Spokes cluster or the Cone(C) region 
is higher ($\Delta \log \xi = +0.43 \pm 0.12$
in Spokes cluster and $\Delta \log \xi = +0.24 \pm 0.15$ in Cone(C) in the mass 
interval of $\log m =$ +0.5 -- -0.5) than the average value of four 
active SFRs. And Cone(H) is slightly lower ($\Delta \log \xi = -0.31 \pm 
0.18$ in the same mass interval). The IMF of the Spokes cluster shows a slight 
enhancement at about 2.5 M$_\odot$ above the mean (about 2 $\sigma$ level). 
The surface density of the Halo region is very low ($\Delta \log \xi = -0.76 
\pm 0.17$) and lacks massive stars. The low surface density may be due 
partly to the incompleteness of the membership selection in the Halo region.
The IMF of the
Field region is far lower. This is due
partly to the incompleteness of membership selection but largely
to the huge area observed relative to the small SFRs spreading over the Field 
region.

\subsection{The IMF from Color-Magnitude Diagrams}

The IMF of open clusters is derived from their CMDs.
But for young open clusters where the distribution of reddening material
is very inhomogeneous, there are many limitations, especially for
low-mass PMS stars, because there is practically no ways to determine
the reddening of individual stars from photometric data alone. In addition,
there is the uncertainty in PMS evolution models. We derived 
the IMF of stars in the PMS locus of NGC 2264 \citep{sbc08} by applying 
the mean reddening. In this case we could not take the heavily embedded stars 
and BMS stars into account. We derived the IMF of NGC 2264 for two values
of reddening - 
the mean reddening (E($B-V$) = 0.07 mag - \citet{sbl97,psbk00}) 
determined from early type stars in NGC 2264, and the median A$_V$ from
SED fitter for each region. As already checked in \citet{sbc04} a small
difference in reddening does not greatly affect the shape of the IMF, i.e.
the more important factor is the inclusion or exclusion of embedded stars and
BMS stars. As PMS stars in the PMS locus are mostly less reddened, a
small value of E($B-V$) is more appropriate. The IMF determined from
the stars in the PMS locus is shown in Figure \ref{cmdimf}. The IMF
presented here is the combined IMF of the reference region.
The only massive star (M $>$ 10 M$_\odot$) missed in the derived IMF is NGC 
2264 IRS1, an embedded flat SED object with 13.5 M$_\odot$ from SED fitter.
The shaded area represents the mass range where membership selection
is largely incomplete as mentioned before.

\placefigure{cmdimf}

The IMF of NGC 2264 is very similar to that of the Pleiades \citep{mbsc03}
down to $\log m \approx -1.2$ where the membership selection is incomplete
even in the reference region of NGC 2264. But there is some scatter for
massive stars (m $\geq $ 2 M$_\odot$). This is related to the fluctuations
due to the small number of massive stars in the cluster. The IMF of NGC 2264
shows a slight deficiency in massive stars relative to the IMF of \citet{kro02}.
We could not find a similar peak in the IMF at about 2 M$_\odot$.
The slope of the IMF is $\Gamma = 
-1.65 \pm 0.14$ for $\log m > 0.5$. This slope is much steeper than that
of \citet{ees55} or \citet{kro02}, but is very similar to that of 
\citet{jms86,hhcj99}. In this case the surface density in the Spokes cluster 
is very similar to that of S Mon because many members of the Spokes cluster 
are heavily embedded.
The surface density of Cone(C) is slightly elevated, and that of Cone(H)
is slightly lower.

\section{YOUNG BROWN DWARF CANDIDATES}

One of main purposes of this study was to select the YBDCs in NGC 2264. 
NGC 2264 is one of the nearest young open clusters and 
the foreground extinction is practically zero. The YBDCs in NGC 2264 
are therefore still very bright and their photometric characteristics
are very similar to those of young low-mass stars. \citet{kbm05} tried
to find the YBDCs in NGC 2264 using deep $Iz$ photometry with CFH12K. 
Although they tried to find YBDCs in various color-color diagrams and
CMDs and through the help of recent model atmospheres, they were unable to
find YBDCs confidently due to a lack of reliable membership selection criteria.
They presented eight probable YBDCs and two additional candidates.

We selected YBDCs with the following criteria.

(1) the mass from CMDs is smaller than 0.075 M$_\odot$. As \citet{sdf00}
published PMS evolution models only for m$_{PMS} \geq$ 0.1 M$_\odot$,
we have to use the PMS evolution models by \citet{bcah98} for the selection of 
YBDCs. But as it is well known that the theoretical colors of \citet{bcah98}
are not well matched for low-mass stars, we have transformed their 
model parameters (T$_{eff}$ and L$_{Bol}$) to $V-I$ or $R-I$ using
the empirical color-temperature and color-BC relations by \citet{msb91}.
Still the theoretical evolutionary tracks by \citet{bcah98} do not
match well with those of \citet{sdf00}, but it is now possible to extrapolate
the tracks to the red color regime by following the 0.1 M$_\odot$
track of \citet{sdf00}.

(2) Stars with one or more membership criteria such as H$\alpha$ emission,
X-ray emission, or IR excess (Class I, II, or III objects).

(3) The mass from SED fitter if the star is detected with the {\it Spitzer}
should be 0.1 M$_\odot$ which is the lowest mass from SED fitter.

Using these criteria, we have selected the 79 YBDCs presented in Table
\ref{bdctab}. The CMDs of these YBCDs are shown in Figure \ref{bdcmd}.
The PMS evolutionary tracks are superimposed in the figure. IR excess
YBDCs are relatively bright due to the limited exposure time used in 
the {\it Spitzer} observations. In addition, X-ray emission cannot identify
YBDCs fainter than $I \approx 19$ and while H$\alpha$ photometry reaches
down to fainter limits, the selection probability is somewhat low
(about 30\% - see \citet{sbc04}).

Three of the YBDCs in Table \ref{bdctab} (C33215 = 2MASS J06405674+0938101, C35463
= J06410604+0949232, and C20186 = J06401789+0941546) are in common with the YBDCs 
listed by \citet{kbm05}. Another object (2MASS J06402759+0945464) could be 
a YBCD because the star is within the PMS locus of both CMDs and $R-$H$\alpha$
is very similar to that of weak line T Tauri stars with X-ray emission.
The other 6 YBDCs in \citet{kbm05} may not be members of NGC 2264 -
Three objects (2MASS J06395722+0941011,
J06400642+0944197, J06411281+0945529) are fainter than the lower boundary of
PMS locus of NGC 2264 \citep{sbc08}. Two (J06401053+0939557, J06413132+0935120)
are within the PMS locus in the ($I$, $V-I$) diagram, but are below the lower
boundary of the PMS locus in the ($I$, $R-I$) diagram. In addition the 
$R-$H$\alpha$ colors of these two objects are very similar to those of 
reddened background stars. And 2MASS J06404873+0939017 is detected only in $I$. 

\placefigure{bdcmd}
\placetable{bdctab}

\section{Discussion}

\subsection{The Effect of Binarity on the IMF}

The fraction of binary or multiple systems and their characteristics are 
important
constraints for the studies of stellar formation and evolution \citep{ajb07}.
Several studies on the frequency of multiple systems and their properties were 
made especially for very low-mass stars (see \citet{sd03} and references
therein), but their initial values are still not well known. The frequency and
mass ratio in multiple systems may affect the actual shape of the IMF.

In general the IMF derived from young open clusters is just the IMF of single 
stars and primaries of multiple systems. It is very important to check the 
effect of multiple systems on the IMF. We constructed a model cluster of 1,000
primaries plus single stars with a Monte Carlo method. The IMF of NGC 2264 
in Figure \ref{cmdimf} is adopted as an input IMF. The frequency of binary
as a function of primary mass was adopted from the results of two-step dynamical 
model of \citet{sd03} which reproduce well the observations. In addition the
secondary to primary mass ratio as a function of spectral type in Figure 4 
of \citet{sd03} was converted to a function of primary mass and used in
the simulation. We assume that the binary fraction of massive stars
(m $\geq$ 10 M$_\odot$) was 90\%. As there is nearly no information
on the fraction and mass ratio in multiple systems of triple or more, 
only single stars and binary systems were taken into account.

A total of 355 binary systems and 645 single stars were generated (see Figure
\ref{binary}). The massive
part of the mass function is dominated by the primaries in binary systems, while
the low-mass part is due to single stars. The mass function of secondaries
is very similar to the IMF of all stars. The shape and slope at the massive
part of the IMF of all stars is very similar to those of the input IMF.
In addition, the peak of the IMF occurs at $\log m \approx$ -0.6, which is
the same value as the input IMF. Such similarities are due to the facts
that the binary fraction of very low-mass stars and BDs is nearly zero and
that of massive stars is very high. In addition the mass ratio of very low-mass 
stars and BDs is relatively high, but that of massive stars is preferentially a
smaller value.

\placefigure{binary}

If we change the fraction of binary systems of very low-mass stars and BDs
 from 0\% to 10 \%, the fraction of binary systems increases from 
35.5 \% to 37.1 \% (371 binary systems were generated in this case). 
Although the mass of the IMF peak does not change, the slope of the IMF of the
massive part becomes steeper and the IMF of very low-mass and BD regime stars 
increases.

We can conclude that if the binary fraction and mass ratio distribution are 
very similar to those adopted here, the shape and peak of the IMF
do not change very much. But the fraction of multiple systems and the mass
ratio distribution can affect strongly the IMF of all stars. And therefore
it is very important to investigate the frequency and mass ratio distribution
of multiple systems for several open clusters with different ages.

\subsection{YSO Mass Function of M17 SWex}

Recently \citet{pw10} studied the YSO mass function (YMF) of the embedded 
massive 
star forming region M17 SWex. They used SED fitter to estimate the mass of YSOs 
in M17 SWex and found the YMF to be much steeper than that of \citet{ees55}.
Their YMF showed a peak at m $\approx 3 M_\odot$. The location of the YMF peak is
very similar to the peak in the IMF in Figure \ref{sedimf}. The prominent 
peak in the YMF of M17 SWex may be due to the combination of two effects
- the completeness of the PMS membership selection in 
M17 SWex and the effect mentioned in \S 4.1. 

If we attribute this peak to an artifact from SED fitter, 
then the slope of the YMF of M17 SWex may be somewhat shallower. 
In addition, if we take account of the fact that there are no known
very massive young stars (e.g. M $\gtrsim 30 M_\odot$) with disks,
the most massive stars in the region may have been missed and 
the slope of the YMF of M17 SWex could be normal.

\section{SUMMARY AND CONCLUSION}

We have studied the young open cluster NGC 2264 using all available photometric
data from optical to mid-IR. We performed SED fitting using the $\chi^2$ 
minimizing SED fitting tool by \citet{tpr07} for more than 1,000 stars
detected in all {\it Spitzer} IRAC channels with high quality ($\epsilon_{
IRAC, total} \leq 0.25$ mag) mid-IR data. We estimated the mass and age of
member stars in NGC 2264 and constructed the star formation history and 
IMF of the cluster. The results obtained are summarized as follows.

(1) The parameters relating to the central stars from SED fitter and other
methods are compared. The temperature from SED fitter is, in many cases,
higher than that expected from the spectral type of the star. And the mass of
central stars shows a conspicuous clump at $\log m \approx$ 0.2 -- 0.3 and
-0.2 -- -0.4, which causes anomalous peaks in the IMF from SED fitter.
The age of central stars is, on the other hand, smaller for sub-solar mass
stars, but is larger for massive stars ($m >$ 1 M$_\odot$).

(2) The parameters relating to disks or envelopes of individual stars in 
NGC 2264 were examined and found to be very similar to those in 
\citet{tpr06}. The properties of individual stars showed no 
significant differences among the SFRs in NGC 2264.
As the maximum difference in median age among SFRs is smaller than the age
spread in an SFR, the large scatter of parameters is due mainly to their age 
spread.

(3) The median values of the physical and structural parameters showed an evident
difference among SFRs in NGC 2264 that was strongly related to the median
age of the SFRs. Such a difference among SFRs gives an approximated view of 
several young open clusters with different ages.

(4) The cumulative age distribution of stars showed distinct
differences among SFRs. The median age of SFRs implies that star
formation in NGC 2264 occurred sequentially - star formation
started at the surface of the Monocerotis giant molecular cloud and propagated
into the cloud, i.e. the stars in the Halo and Field region are the oldest,
the stars in S Mon and Cone(H) are intermediate while those in Cone(C) are younger,
and those in the Spokes cluster are the youngest. The star formation history seen 
in NGC 2264 and the fact that the first star formation occurred in the 
low density region (Halo and Field) of NGC 2264 imply that star 
formation in NGC 2264 was triggered externally.

(5) The IMF using the stellar mass from SED fitter shows a prominent peak
at m $\approx$ 2M$_\odot$. As such a peak cannot be found in the IMF of
other open clusters or field stars in the Solar neighborhood, the peak in
the IMF from SED fitter may be an artifact of the SED fitting tool. The IMF
from CMDs is well consistent with the IMF of the Pleiades.

(6) The slope of the IMF of massive stars ($\log m \geq 0.5$) is -1.7 
$\pm$ 0.1. This value is somewhat steeper than the standard Salpeter-Kroupa
IMF \citep{kro02}, but is consistent with that of \citet{jms86}.

(7) We selected 79 young brown dwarf candidates from the PMS loci in the CMDs
of NGC 2264 that showed either H$\alpha$
emission, X-ray emission, or IR excess emission, and with implied masses
smaller than 0.075 M$_\odot$.

\acknowledgments 
The authors would like to thank the anonymous referee for helpful comments.
H.S. acknowledges the support of the National Research Foundation of Korea
(NRF) to the Astrophysical Research Center for
the Structure and Evolution of the Cosmos (ARCSEC$''$) at Sejong University
(NRF No. 2009-0062865).

\clearpage

\clearpage

\clearpage

\clearpage

\clearpage
\begin{figure}
\epsscale{1.0}
\plottwo{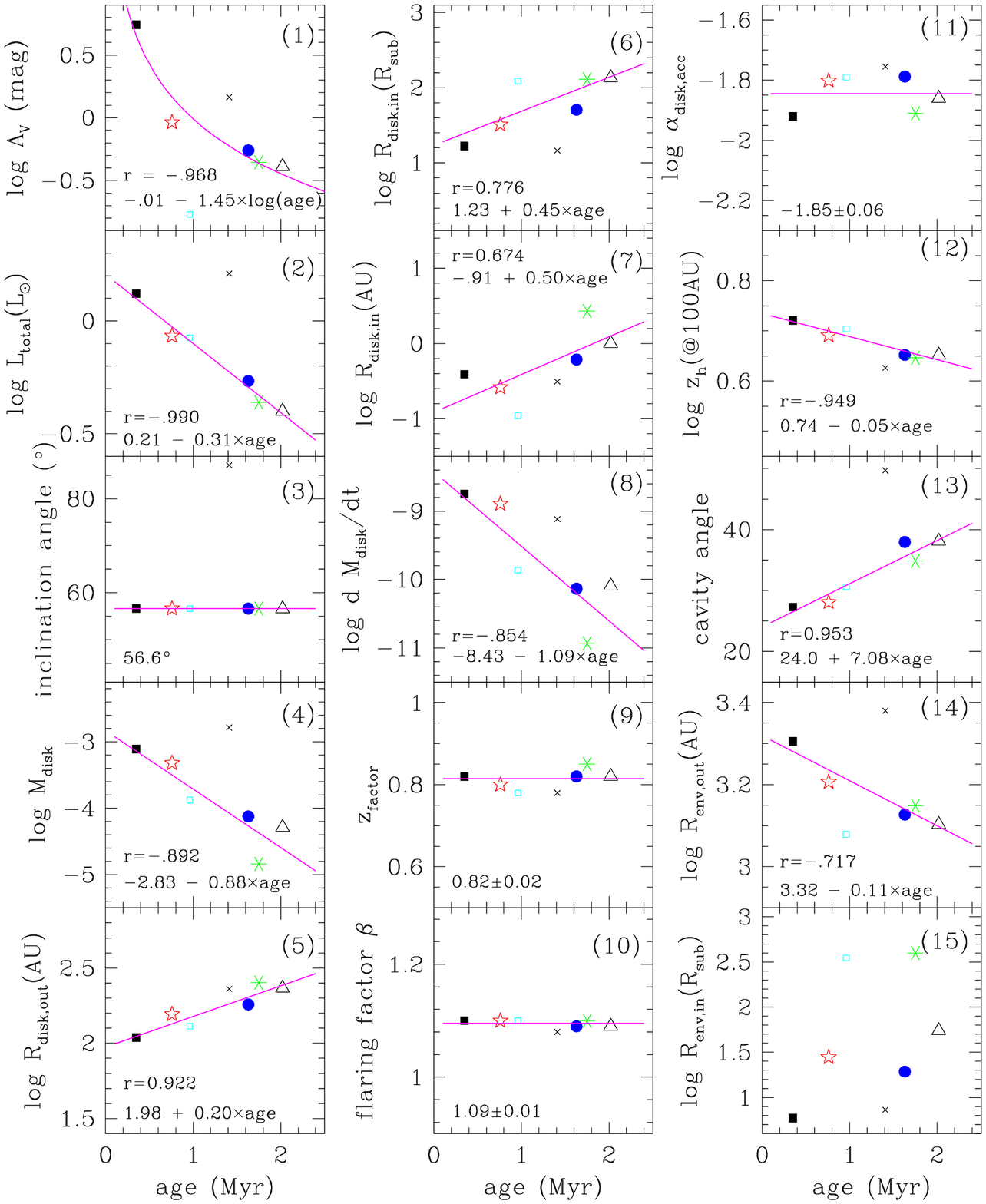}{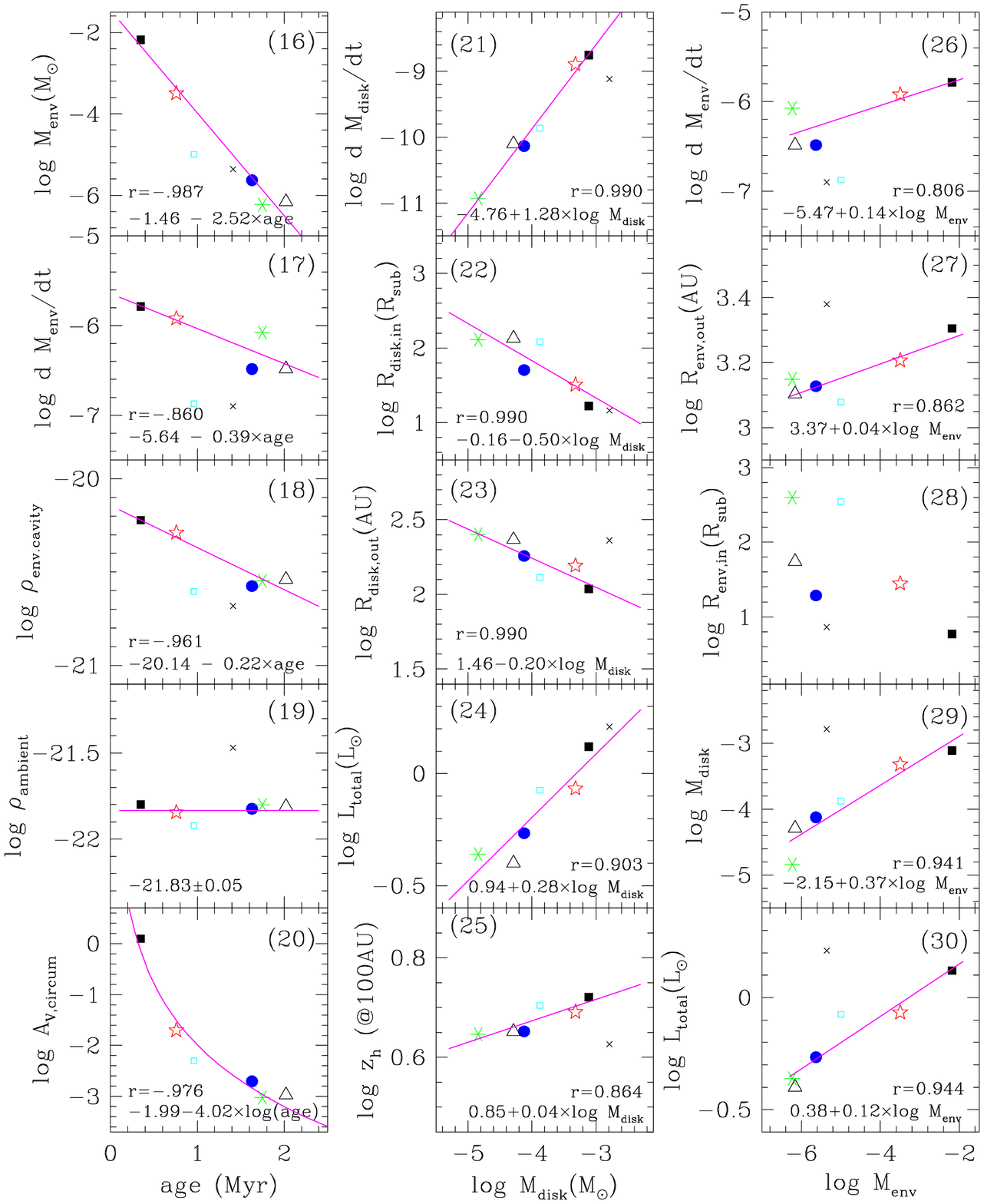}
\caption{Results from SED fitting. To see the evolutionary trend of
YSO parameters, we show the median value of young stars in a given
SFR. A blue dot, black square, red star, green asterisk, triangle,
small open square and small cross represents the median value of young stars
in S Mon, Spokes cluster, Cone(C), Cone(H), Halo, Field, and BMS stars, 
respectively. The actual scattering of individual stars in a given 
SFR is very large and therefore in most cases, it is very 
difficult to see the evolutionary sequence for the region. In addition,
the relations for envelopes are only for small samples of stars with
envelopes in a given SFR.  See text for details. \label{sed_median} }
\end{figure}

\clearpage
\begin{figure}
\epsscale{0.8}
\plotone{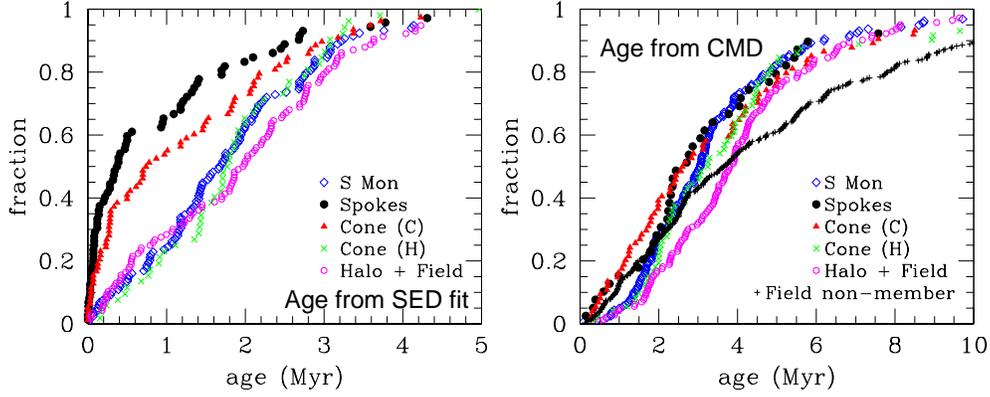}
\caption{Star formation history of NGC 2264. (Left panel) \ Age of PMS stars from
SED fitter. (Right panel) Ages from CMDs. We assumed E($B-V$) = 0.07.  
To avoid unrealistic ages from PMS evolution models, we limit the ages to
PMS stars with m = 1 -- 0.2 M$_\odot$. \label{figsfh} }
\end{figure}

\begin{figure}
\epsscale{0.8}
\plotone{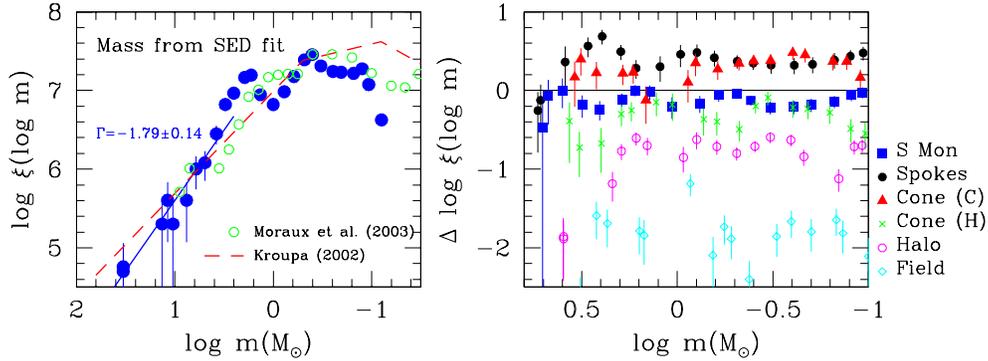}
\caption{The IMF of NGC 2264 from {\it Spitzer} observations. 
(Left panel) The IMF of stars in the S MON and CONE regions (large dots) of 
\citet{sbc08} where membership selection is nearly complete down to 0.25 
M$_\odot$. The IMF of \citet{mbsc03} for the Pleiades (open circle) and 
the IMF \citet{kro02} (dashed line) are superimposed for comparison. 
(Right panel) Differences in the IMF of each region relative to the IMF
in the left panel.  \label{sedimf} }
\end{figure}

\begin{figure}
\epsscale{0.8}
\plotone{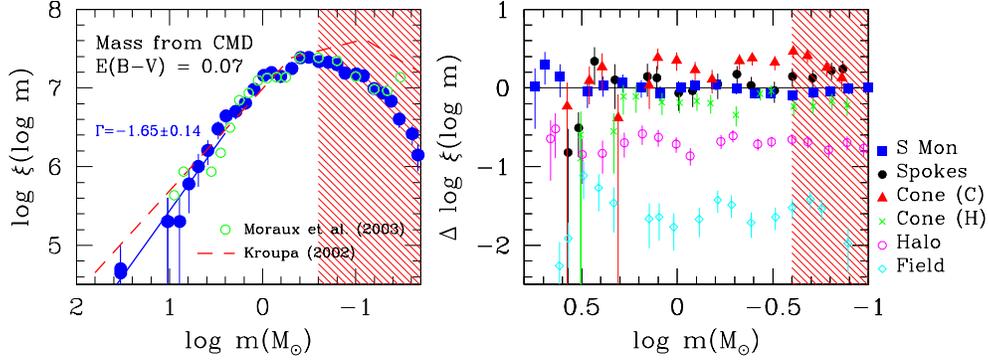}
\caption{The IMF of PMS stars in the PMS locus of NGC 2264 (large dots). 
We assumed E($B-V$) = 0.07. (Left panel) The IMF of stars in the S MON and CONE 
regions of \citet{sbc08} where membership selection is nearly complete down 
to 0.25 M$_\odot$.  (Right panel) Differences in the IMF of each region 
relative to the IMF in the left panel.  \label{cmdimf} }
\end{figure}

\begin{figure}
\epsscale{0.8}
\plotone{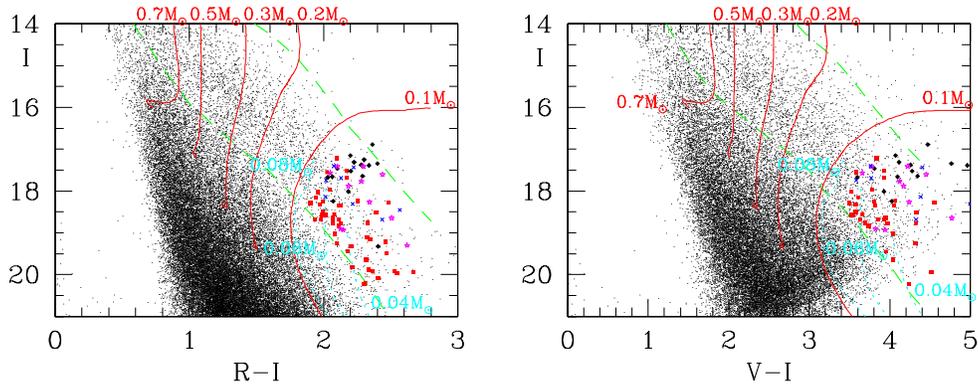}
\caption{Color-Magnitude diagrams of young brown dwarf candidates in NGC 2264.
Small dots are all stars in the field of NGC 2264 \citep{sbc08}. 
The $R-I$ colors of very red stars were corrected using the prescription 
described in Appendix of \citet{sbc08}. The thick dashed lines represent 
the locus of PMS stars in NGC 2264. Thin solid and dotted lines show 
the PMS evolution tracks from \citet{sdf00} and \citet{bcah98}, respectively. 
The empirical color-temperature
and color-BC relations for very red stars from \citet{msb91} were used.
\label{bdcmd} }
\end{figure}

\clearpage
\begin{figure}
\epsscale{0.5}
\plotone{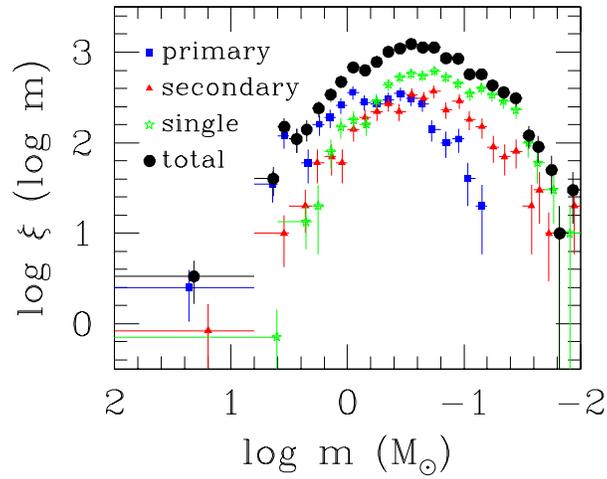}
\caption{The effect of binarity on the IMF. A Monte Carlo simulation was
performed to construct a model cluster of 1,000 primary plus single stars.
The input IMF is the IMF of NGC 2264 in Figure \ref{cmdimf}. The binary
fraction and the mass ratio between primary and secondary as a function 
of primary mass is from \citet{sd03}. Square, triangle, and star symbol
represent, respectively, the IMF of primaries, secondaries, and single stars,
while large dots stand for the IMF of all stars including the secondaries of
binary systems. \label{binary} }
\end{figure}

\clearpage
\begin{deluxetable}{c|cccccc|c}
\tablecolumns{8}
\tabletypesize{\scriptsize}
\tablecaption{Results from SED fitting\tablenotemark{a}}
\tablewidth{0pt}
\rotate
\tablehead{
\colhead{parameter} & \colhead{S Mon} & \colhead{Spokes cluster} & \colhead{Cone(C)} & 
\colhead{Cone(H)} & \colhead{Halo} & \colhead{Field} & \colhead{BMS stars\tablenotemark{b} }}
\startdata
f$_{Stage ~0/1}$ & 0.19 & 0.52 & 0.39 & 0.15 & 0.18 & 0.18 & 0.10 \\
f$_{Stage ~2}$ & 0.54 & 0.34 & 0.42 & 0.52 & 0.53 & 0.58 & 0.85 \\
age (Myr)\tablenotemark{c} & 1.63 $_{0.27}^{3.27}$ (120) & 
          0.35 $_{0.02}^{2.55}$ (78) & 0.76 $_{0.04}^{3.16}$ (82) & 
          1.75 $_{0.40}^{3.13}$ (58) & 2.02 $_{0.32}^{3.73}$ (102) & 
          0.96 $_{0.29}^{3.41}$ (14) & 1.41 $_{0.40}^{8.34}$ (19) \\
A$_{V}$ & 0.52 $_{0.02}^{2.81}$ (257) & 5.74 $_{0.29}^{40.5}$ (176) & 
          0.92 $_{0.04}^{31.2}$ (145) &  0.44 $_{0.01}^{3.13}$ (100) & 
          0.41 $_{0.01}^{4.01}$ (177) & 0.17 $_{0.02}^{2.11}$ (33) & 
          1.46 $_{0.03}^{3.66}$ \\
$\log d(kpc)$ & -0.09 $_{-.15}^{0.00}$ (257) & -0.11 $_{-.15}^{-.02}$ (176) & 
         -0.11 $_{-.15}^{-.02}$ (145) & -0.09 $_{-.15}^{0.00}$ (100) & 
         -0.09 $_{-.15}^{0.00}$ (177) & -0.11 $_{-.15}^{-.02}$ (33) &
         -0.10 $_{-.15}^{0.00}$ (40) \\
$\log L_{total}$ (L$_\odot$) & -0.27 $_{-.92}^{1.02}$ (257) & 
          0.12 $_{-.59}^{1.52}$ (176) & -0.07 $_{-.67}^{1.18}$ (145) & 
         -0.36 $_{-.79}^{0.96}$ (100) & -0.40 $_{-.86}^{0.99}$ (177) & 
         -0.07 $_{-.92}^{1.70}$ (33) & 0.21 $_{-.54}^{1.40}$ (40) \\
disk & & & & & & & \\
inclination ($^\circ$)& 56.6 $_{18.2}^{81.4}$ (257) & 56.6 $_{18.2}^{81.4}$ (176) & 
          56.6 $_{18.2}^{75.5}$ (145) & 56.6 $_{18.2}^{81.4}$ (100) & 
          56.6 $_{18.2}^{81.4}$ (177) & 56.6 $_{18.2}^{86.0}$ (33) &
          87.1 $_{68.2}^{87.1}$ (40) \\
$\log$ M$_d$ (M$_\odot$) & -4.12 $_{-7.87}^{-2.07}$ (257) & 
         -3.11 $_{-6.73}^{-1.81}$ (176) & -3.32 $_{-7.36}^{-1.89}$ (145) & 
         -4.84 $_{-7.97}^{-2.24}$ (100) & -4.29 $_{-7.84}^{-2.21}$ (177) & 
         -3.88 $_{-7.43}^{-2.14}$ (33) & -2.79 $_{-4.94}^{-1.58}$ (40) \\
$\log R_{d,out}$ (AU) & 2.26 $_{1.69}^{3.10}$ (257) & 2.04 $_{1.33}^{3.02}$ (176) & 
          2.19 $_{1.50}^{3.01}$ (145) & 2.40 $_{1.64}^{3.31}$ (100) & 
          2.37 $_{1.83}^{3.26}$ (177) & 2.11 $_{1.66}^{2.99}$ (33) &
          2.36 $_{2.01}^{3.19}$ (40) \\
$\log R_{d,in}$ (R$_{sub}$)\tablenotemark{d} & 1.71 $_{0.45}^{2.76}$ (181) & 
          1.23 $_{0.26}^{2.70}$ (111) & 1.51 $_{0.32}^{2.78}$ (82) & 
          2.11 $_{0.67}^{2.75}$ (72) & 2.13 $_{0.64}^{2.94}$ (126) & 
          2.09 $_{0.35}^{2.72}$ (14) & 1.16 $_{0.32}^{2.65}$ (22) \\
$\log R_{d,in}$ (AU) & -0.21 $_{-1.45}^{1.63}$ (257) & -0.41 $_{-1.29}^{1.40}$ (176) & 
         -0.58 $_{-1.44}^{1.58}$ (145) & 0.43 $_{-1.48}^{1.53}$ (100) & 
          0.00 $_{-1.48}^{1.68}$ (177) & -0.96 $_{-1.53}^{1.58}$ (33) &
         -0.51 $_{-1.25}^{1.18}$ (40) \\
$\log \dot{M_{d}}$ (M$_\odot /$ yr) & -10.13 $_{-13.52}^{-7.21}$ (257) & 
         -8.75 $_{-13.07}^{-6.64}$ (176) & -8.89 $_{-13.34}^{-6.80}$ (145) & 
         -10.93 $_{-14.36}^{-7.49}$ (100) & -10.10 $_{-13.67}^{-7.81}$ (177) & 
         -9.86 $_{-13.30}^{-7.52}$ (33) & -9.12 $_{-11.33}^{-7.13}$ (40) \\
scale height factor & 0.82 $_{0.64}^{0.97}$ (257) & 0.82 $_{0.64}^{0.96}$ (176) & 
          0.80 $_{0.63}^{0.97}$ (145) & 0.85 $_{0.65}^{0.97}$ (100) & 
          0.82 $_{0.63}^{0.96}$ (177) & 0.78 $_{0.63}^{1.00}$ (33) &
          0.78 $_{0.59}^{0.96}$ (40) \\
flaring power $\beta$ & 1.09 $_{1.02}^{1.16}$ (257) & 1.10 $_{1.02}^{1.17}$ (176) & 
          1.10 $_{1.03}^{1.16}$ (145) & 1.10 $_{1.02}^{1.16}$ (100) & 
          1.09 $_{1.02}^{1.16}$ (177) & 1.10 $_{1.04}^{1.15}$ (33) &
          1.08 $_{1.03}^{1.14}$ (40) \\
disk accetion $\alpha$ & -1.79 $_{-2.67}^{-1.13}$ (257) & 
         -1.92 $_{-2.85}^{-1.14}$ (176) & -1.80 $_{-2.77}^{_1.17}$ (145) & 
         -1.91 $_{-2.82}^{-1.13}$ (100) & -1.86 $_{-2.80}^{-1.22}$ (177) & 
         -1.79 $_{-2.81}^{-1.17}$ (33) & -1.76 $_{-2.86}^{-1.05}$ (40) \\
$h$ (at 100AU) & 4.49 $_{2.34}^{8.88}$ (257) & 5.26 $_{2.79}^{11.4}$ (176) & 
          4.91 $_{2.70}^{10.6}$ (145) & 4.43 $_{2.76}^{8.31}$ (100) & 
          4.49 $_{2.33}^{9.76}$ (177) & 5.06 $_{3.31}^{8.84}$ (33) &
          4.23 $_{2.22}^{7.68}$ (40) \\
envelope & & & & & & & \\
cavity angle ($\circ$) & 37.9 $_{19.9}^{52.6}$ (90) & 27.3 $_{9.0}^{45.4}$ (117) & 
          28.1 $_{10.7}^{49.6}$ (81) & 34.8 $_{16.4}^{51.8}$ (26) & 
          38.1 $_{18.1}^{50.5}$ (56) & 30.6 $_{21.3}^{46.3}$ (14)  &
          49.7 $_{31.9}^{56.9}$ (16) \\
$\log R_{env,out}$ (AU) & 3.13 $_{3.02}^{3.63}$ (90) & 3.31 $_{3.07}^{3.78}$ (117) & 
          3.21 $_{3.03}^{3.64}$ (81) & 3.15 $_{3.01}^{3.50}$ (26) & 
          3.10 $_{3.01}^{3.52}$ (56) & 3.08 $_{3.01}^{3.37}$ (14) &
          3.38 $_{3.03}^{3.87}$ (16) \\
$\log R_{env,in}$ (R$_{sub}$)\tablenotemark{d} & 1.29 $_{0.39}^{2.71}$ (63) & 
          0.77 $_{0.25}^{2.60}$ (79) & 1.45 $_{0.38}^{2.80}$ (48) & 
          2.60 $_{1.72}^{2.92}$ (19) & 1.74 $_{0.60}^{2.70}$ (37) & 
          2.54 $_{1.59}^{2.72}$ (5) & 0.86 $_{0.29}^{1.63}$ (10) \\
$\log$ M$_{env}$ (M$_\odot$) & -5.63 $_{-8.48}^{-1.76}$ (257) & 
         -2.18 $_{-7.81}^{-0.70}$ (176) & -3.50 $_{-8.17}^{-1.11}$ (145) & 
         -6.23 $_{-8.56}^{-2.05}$ (100) & -6.16 $_{-8.36}^{-2.16}$ (177) & 
         -5.00 $_{-8.20}^{-1.91}$ (33) & -5.36 $_{-8.11}^{-2.35}$ (40) \\
$\log{\dot{M_{env}}}$ (M$_\odot /$ yr) & -6.48 $_{-8.57}^{-5.12}$ (90) & 
         -5.79 $_{-6.68}^{-4.65}$ (117) & -5.92 $_{7.88}^{-4.99}$ (81) & 
         -6.08 $_{-8.10}^{-5.23}$ (26) & -6.48 $_{-8.32}^{-5.13}$ (56) & 
         -6.87 $_{-8.13}^{-5.46}$ (14) & -6.90 $_{-7.60}^{-6.09}$ (16) \\
$\log \rho_{cavity}$ (g cm$^{-3}$) & -20.6 $_{-21.2}^{-19.8}$ (90) &
         -20.2 $_{-20.8}^{-19.5}$ (117) & -20.3 $_{-21.1}^{-19.5}$ (81) &
         -20.5 $_{-21.1}^{-19.7}$ (56) & -20.6 $_{-20.9}^{-20.2}$  (14) &
         -20.7 $_{-21.1}^{-20.4}$ (16) \\
ambient & & & & & & & \\
$\log \rho_{amb}$ (g cm$^{-3}$) & -21.8 $_{-22.0}^{-21.1}$ (257) &
         -21.8 $_{-22.0}^{-21.1}$ (176) & -21.8 $_{-22.0}^{-21.2}$ (145) & 
         -21.8 $_{-22.0}^{-21.2}$ (100) & -21.8 $_{-22.0}^{-21.1}$ (177) &
         -21.9 $_{22.0}^{21.2}$ (33) & -21.5 $_{22.0}^{-21.0}$ (40) \\
$\log A_{V,circum}$ & -2.70 $_{-4.07}^{0.73}$ (257) & 0.10 $_{-3.44}^{1.74}$ (176) & 
         -1.70 $_{-3.82}^{1.52}$ (145) & -3.03 $_{-4.09}^{0.37}$ (100) & 
         -2.97 $_{-4.10}^{0.76}$ (177) & -2.30 $_{-4.01}^{2.43}$ (33) &
         2.19 $_{-2.78}^{4.17}$ (40) \\
\enddata
\tablenotetext{a}{The data given are median, 10\%, 90\%, and number used in the calculation.}
\tablenotetext{b}{We choose the $\chi_{min}$ model as the best model for BMS stars.}
\tablenotetext{d}{Due to the uncertainty in the age of massive PMS stars, the mass limit used here is m = 0.2 -- 1.0 M$_\odot$.}
\tablenotetext{d}{We neglect the data whose value is 1.00 $R_{sub}$.}
\label{sedtab}
\end{deluxetable}

\clearpage
\begin{deluxetable}{cc|cccc|ccccc}
\tablecolumns{11}
\tabletypesize{\scriptsize}
\tablecaption{Ages and K-S Test Results}
\tablewidth{0pt}
\rotate
\tablehead{
\colhead{method} & \colhead{SFR} & \colhead{N$_{\rm star}$\tablenotemark{a}} & 
\colhead{$\tau_{median}$ (Myr)\tablenotemark{a}} & 
\colhead{$\tau_{10\%}$ (Myr)\tablenotemark{a}} & \colhead{$\tau_{90\%}$ (Myr)\tablenotemark{a}} & 
\multicolumn{5}{c}{Probability from Kolmogorov-Smirnov Test} }

\startdata
reference & & & & & & S Mon & Spokes cluster & Cone(C) & Cone(H) & Halo \\ \hline
& S Mon & 120 & 1.63 & 0.27 & 3.27 & \nodata & \nodata & \nodata & \nodata & \nodata \\
& Spokes cluster & 78 & 0.35 & 0.02 & 2.55 & 1.65$\times 10^{-8}$ & \nodata & \nodata & \nodata & \nodata\\
SED & Cone(C) & 82 & 0.76 & 0.04 & 3.16 & 9.54$\times 10^{-5}$ & 0.148 & \nodata & \nodata & \nodata\\
& Cone(H) & 58 & 1.75 & 0.40 & 3.13 & 0.858 & 1.04$\times 10^{-6}$ & 5.97$\times 10^{-4}$ & \nodata & \nodata\\
& Halo & 102 & 2.02 & 0.32 & 3.73 & 0.066 & 3.40$\times 10^{-8}$ & 1.99$\times 10^{-4}$ & 0.281 & \nodata \\
& Field & 14 & 0.96 & 0.29 & 3.41 & 0.091 & 6.60$\times 10^{-3}$ & 0.065 & 0.090 & 0.263 \\ \hline
& S Mon & 159 & 3.05 & 1.41 & 6.20 & \nodata & \nodata & \nodata & \nodata & \nodata  \\
& Spokes cluster & 39 & 2.71 & 0.72 & 6.15 & 0.776 & \nodata & \nodata & \nodata & \nodata  \\
& Cone(C) & 74 & 2.64 & 0.69 & 7.54 & 0.140 & 0.786 & \nodata & \nodata & \nodata \\
CMD & Cone(H) & 72 & 3.24 & 1.67 & 8.75 & 0.372 & 0.824 & 0.110 & \nodata & \nodata \\
& Halo & 110 & 3.80 & 1.67 & 7.45 & 4.05$\times 10^{-4}$ & 0.266 & 1.51$\times 10^{-2}$ & 0.202 & \nodata \\
& Field & 33 & 3.91 & 2.00 & 6.50 & 2.34$\times 10^{-2}$ & 0.060 & 2.82$\times 10^{-2}$ & 0.285 & 0.922 \\
& Field nm\tablenotemark{b} & 307 & 3.66\tablenotemark{b} & 0.91\tablenotemark{b} & 10.14\tablenotemark{b} & 1.04$\times 10^{-4}$ & 0.117 & 5.88$\times 10^{-2}$ & 3.05$\times 10^{-3}$ & 2.54$\times 10^{-2}$ \\

\enddata
\tablenotetext{a}{Due to the uncertainty in the age of massive PMS stars, the mass limit used here is m = 0.2 -- 1.0 M$_\odot$.}
\tablenotetext{b}{``Field nm'' represents the stars in the Field region. 
They are in the PMS locus of NGC 2264 CMD, but do not show any appreciable 
H$\alpha$ emission or IR excess. And therefore their ``age'' has meaningless
because they are not the member of NGC 2264. But to check the reliability
of Kolmogorov-Smirnov test, we estimate the age of these non-members from
the position in CMDs.}
\label{agetab}
\end{deluxetable}

\begin{deluxetable}{ccccccc}
\tablecolumns{10}
\tabletypesize{\scriptsize}
\tablecaption{Young Brown Dwarf Candidates}
\tablewidth{0pt}
\tablehead{
\colhead{Sung et al. (2008)} & \colhead{$I$} & \colhead{$R-I$} & \colhead{$V-I$} & \colhead{H$\alpha$ or X-ray} & \colhead{2MASS} & \colhead{YSO Class} }

\startdata
C3200 & 18.290& 1.901& 3.500 &H$\alpha$      &06393962+0940442&     \\
C18770 & 18.669& 1.977& 3.636 &H$\alpha$      &06401412+0934091&     \\
C20186 & 18.651& 2.067& 3.946 &H$\alpha$      &06401789+0941546&     \\
C21372 & 17.508& 2.176& 3.925 &               &06402107+0935247& II  \\
C21459 & 19.205& 2.462&\nodata&H$\alpha$      &                &     \\
C21605 & 18.516& 2.394& 4.333 &H$\alpha$      &                &     \\
C22265 & 18.581& 2.016& 3.540 &H$\alpha$      &06402354+0948330&     \\
C23335 & 19.023& 2.085& 4.037 &H$\alpha$      &                &     \\
C24265 & 19.914& 2.365&\nodata&H$\alpha$      &                &     \\
C24405 & 18.554& 2.079& 3.659 &H$\alpha$      &06402937+0941035&     \\
C24851 & 18.011& 2.184& 3.926 &H$\alpha$      &06403063+0954365& II  \\
C24999 & 17.224& 2.096& 3.840 &H$\alpha$      &06403106+0935401&     \\
C25017 & 17.525& 2.202& 4.275 &X+H$\alpha$    &06403110+0949319& II  \\
C28287 & 19.907& 2.527&\nodata&H$\alpha$      &06404043+0911372&     \\
C28403 & 17.739& 2.019& 3.579 &X              &06404078+0934269&     \\
C28559 & 18.685& 2.438& 4.326 &X              &06404119+0931287&     \\
C28717 & 17.566& 2.050& 3.659 &X+H$\alpha$(Sp)&06404164+0931431&     \\
C28798 & 19.147& 2.349&\nodata&H$\alpha$(Sp)  &                &     \\
C29818 & 17.606& 2.439& 4.452 &X+H$\alpha$C   &06404494+0938497&     \\
C30192 & 17.671& 2.171& 3.931 &H$\alpha$(Sp)  &06404610+0947142&     \\
C31370 & 20.197& 2.311&\nodata&H$\alpha$      &                &     \\
C31392 & 17.876& 2.186& 4.169 &X+H$\alpha$(Sp)&                &     \\
C31403 & 17.693& 2.129& 3.768 &X              &06405006+0947064&     \\
C31736 & 17.549& 2.025& 3.581 &x+H$\alpha$    &06405104+0949061&     \\
C31751 & 18.152& 1.943& 3.608 &H$\alpha$      &                &     \\
C31863 & 18.140& 2.013& 4.381 &X              &06405139+0945576&     \\
C32294 & 18.829& 2.116& 3.741 &H$\alpha$C     &06405300+0947377&     \\
C32976 & 19.323& 2.407&\nodata&X              &06405579+0936061& II  \\
C33131 & 18.330& 1.973& 3.579 &H$\alpha$      &06405641+0934442&     \\
C33215 & 19.098& 2.323&\nodata&H$\alpha$      &06405674+0938101&     \\
C33403 & 18.285& 2.481& 4.557 &x+H$\alpha$    &06405747+0935147&     \\
C33533 & 17.462& 2.102& 3.809 &H$\alpha$C     &06405798+1002163& II  \\
C33590 & 17.766& 2.280& 4.122 &X+H$\alpha$C   &                &     \\
C33624 & 17.374& 2.280& 4.256 &X+H$\alpha$    &06405834+0937567& II  \\
C33627 & 20.070& 2.409&\nodata&H$\alpha$      &                &     \\
C33648 & 16.886& 2.363& 4.471 &x              &06405843+0937445& II  \\
C33894 & 17.675& 2.030& 3.596 &H$\alpha$C     &06405941+0945516& II  \\
C33920 & 19.328& 2.207&\nodata&H$\alpha$      &                &     \\
C33939 & 17.641& 2.291& 4.344 &X+H$\alpha$C   &06405962+0936575& II  \\
C34020 & 19.809& 2.214&\nodata&H$\alpha$      &                &     \\
C34101 & 18.344& 2.112& 3.813 &H$\alpha$C     &06410032+0930169&     \\
C34164 & 17.430& 2.224& 4.404 &X              &                &     \\
C34674 & 18.456& 2.569&\nodata&X              &06410258+0936156&     \\
C34695 & 17.415& 2.104& 3.774 &X+H$\alpha$(Sp)&06410266+0950329&     \\
C34875 & 18.497& 1.995& 3.571 &H$\alpha$C     &06410350+0935248&     \\
C34965 & 18.900& 2.124& 3.820 &X+H$\alpha$    &                &     \\
C34989 & 17.661& 2.065& 4.094 &H$\alpha$      &06410389+0919352& II  \\
C35065 & 17.406& 2.296& 4.223 &X+H$\alpha$    &06410417+0934572&     \\
C35418 & 17.311& 2.227& 4.115 &X+H$\alpha$    &06410579+0931072& II  \\
C35463 & 18.263& 2.343& 4.241 &X+H$\alpha$    &06410604+0949232&     \\
C35544 & 20.220& 2.301& 4.236 &H$\alpha$      &                &     \\
C35632 & 18.929& 2.147& 3.939 &X+H$\alpha$C   &06410696+0935557&     \\
C35763 & 18.311& 2.216& 4.994 &X              &06410755+0930004&     \\
C35876 & 17.344& 2.401& 4.765 &H$\alpha$      &06410812+0918151& II  \\
C36048 & 17.386& 2.337& 4.963 &X              &                & II  \\
C36258 & 19.928& 2.673& 4.519 &H$\alpha$      &06411016+0931223&     \\
C36580 & 18.646& 2.508& 4.767 &X+H$\alpha$    &06411180+0931123&     \\
C36630 & 18.242& 2.071& 4.024 &H$\alpha$C     &06411208+0945281& II  \\
C36702 & 17.151& 2.199& 4.069 &               &06411242+0913247& II  \\
C36868 & 17.253& 2.302&\nodata&               &06411326+0935473&II/III\\
C36941 & 18.357& 2.130& 4.047 &H$\alpha$      &06411365+0955370&     \\
C37325 & 19.955& 2.548&\nodata&H$\alpha$      &06411578+0915357&     \\
C37393 & 18.666& 2.111& 3.929 &H$\alpha$C     &06411613+0938166&     \\
C37601 & 19.449& 2.178& 3.953 &H$\alpha$C     &                &     \\
C37602 & 18.017& 1.958& 3.523 &H$\alpha$      &06411714+0943378&     \\
C37611 & 18.672& 1.931& 3.854 &H$\alpha$      &06411714+0929048&     \\
C37654 & 18.232& 2.257& 4.066 &H$\alpha$      &06411736+0931519&     \\
C37667 & 19.301& 2.621&\nodata&X+H$\alpha$    &06411737+0937151&     \\
C37980 & 18.770& 2.325& 4.325 &H$\alpha$C     &06411862+0954074&     \\
C38020 & 19.648& 2.315& 4.042 &H$\alpha$      &                &     \\
C38517 & 19.247& 2.100& 4.313 &H$\alpha$      &                &     \\
C38598 & 17.402& 2.076& 3.762 &X              &06412104+0932417&     \\
C38760 & 18.262& 1.992& 3.806 &H$\alpha$      &06412177+1017003&     \\
C39183 & 18.760& 2.067& 3.700 &H$\alpha$      &                &     \\
C39897 & 17.970& 2.047& 3.780 &H$\alpha$      &06412593+0930260&     \\
C40179 & 18.567& 1.988& 3.528 &H$\alpha$      &06412695+0946003&     \\
C42417 & 19.487& 2.466&\nodata&H$\alpha$C     &                &     \\
C43063 & 18.778& 2.130& 3.976 &H$\alpha$      &06413618+1010481&     \\
C44891 & 19.875& 2.333&\nodata&H$\alpha$      &                &     \\
\enddata
\label{bdctab}
\end{deluxetable}


\begin{thebibliography}{200}

\bibitem[Alves et al.(2007)]{all07} Alves, J., Lombardi, M., \& Lada, C. J.
    2007, \aap, 462, L17
\bibitem[Ballero et al.(2007)]{bkm07} Ballero, S. K., Kroupa, P., \& Matteucci,
    F. 2007, \aap, 467, 117
\bibitem[Baraffe et al.(1998)]{bcah98} Baraffe, I., Chabrier, G., Allard, F.,
    \& Hauschildt, P. H. 1998, \aap, 337, 403
\bibitem[Bartko et al.(2010)]{hb10} Bartko, H. et al. 2010, \apj, 708, 834
\bibitem[Baxter et al.(2009)]{ejb09} Baxter, E. J., Covey, K. R., Muench, A. A.,
    F\"ur\'esz, G., Rebull, L., \& Szentgyorgyi, A. H. 2009, \aj, 138, 963
\bibitem[Bessell(1991)]{msb91} Bessell, M. S. 1991, \aj, 101, 662
\bibitem[Bessell(1995)]{msb95} Bessell, M. S. 1995, in The Bottom of the Main
    Sequence and Beyond, ed. by C. Tinney, (Springer, Berlin) p123
\bibitem[Billot et al.(2010)]{nb10} Billot, N., Noriega-Crespo, A., Carey, S.,
    Guieu, S., Shenoy, S., Paladini, R., \& W. Latter 2010, arXiv:1003.0866
\bibitem[Blaauw(1964)]{ab64} Blaauw, A. 1964, \araa, 2, 213
\bibitem[Bonnell et al.(2001)]{bbcp01} Bonnell, I. A., Bate, M. R., Clarke, C.
    J., \& Pringle, J. E. 2001, \mnras, 323, 785
\bibitem[Burgasser et al.(2007)]{ajb07} Burgasser, A. J., Reid, I. N., Siegler, 
    N., Close, L., Allen, P., Lowrance, P., \& Gizis, J. 2007, Protostars and
    Planets V, eds by B. Reipurthm D. Jewitt, and K. Keil, (Univ of Arizona 
    Press, Tucson), p951
\bibitem[Chabrier(2001)]{cha01} Chabrier, G. 2001, ApJ, 554, 1274 
\bibitem[Chun et al.(2010)]{clsb10} Chun, M.-Y., Lim, B., Sung, H., Besell, M. 
    S., \& Sohn, T. 2010, in preparation
\bibitem[Dahm \& Simon(2005)]{ds05} Dahm, S. E., \& Simon, T. 2005, \aj,
    129, 829
\bibitem[Dib et al.(2010)]{sd10} Dib, S., Shadmehri, M., Padoan, P., 
    Maheswar, G., Ojha, D. K., \& Khajenabi, F. 2010, \mnras, in press
\bibitem[Drilling(1975)]{jsd75} Drilling, J. S. 1975, \aj, 80, 128
\bibitem[Elmegreen(1998)]{elm98} Elmegreen, B. G. 1998, ASP Conf. Ser., 148, 150
\bibitem[Elmegreen \& Scalo(2006)]{es06} Elmegreen, B. G., \& Scalo, J. 2006, 
    \apj, 636, 149
\bibitem[Elmegreen \& Shadmehri(2003)]{es03} Elmegreen, B. G., \& Shadmehri, M.
    2003, \mnras, 338, 817
\bibitem[Espinoza et al.(2009)]{esm09} Espinoza, P., Selman, F. J., Melnick, J.
    2009, \aap, 501, 563
\bibitem[Fazio et al.(2004)]{irac} Fazio, G. G. et al. 2004, \apjs, 154, 10
\bibitem[Flaccomio et al.(2006)]{fl06} Flaccomio, E., Micela, G., \& Sciortino,
    S. 2006, \aap, 455, 903
\bibitem[F\"ur\'esz, G. et al.(2006)]{gf06} F\"ur\'esz, G. et al. 2006, \aj,
    648, 1090
\bibitem[Gies et al.(1996)]{drg97} Gies, D. R. et al. 1997, \apj, 475, L49
\bibitem[Hambly et al.(1999)]{hhcj99} Hambly, N. C., Hodgkin, S. T., 
    Cossburn, M. R., \& Jameson, R. F. 1999, \mnras, 303, 835
\bibitem[Hartmann(2003)]{lh03} Hartmann, L. 2003, \apj, 585, 398
\bibitem[Heiles(1979)]{ch79} Heiles, C. 1979, \apj, 229, 533
\bibitem[Herbig(1962)]{gh62} Herbig, G. 1962, \apj, 135, 736
\bibitem[Hur et al.(2010)]{hsb10} Hur, H., Sung, H., \& Bessell, M. S. 2010,
    in preparation
\bibitem[Kendall et al.(2005)]{kbm05} Kendall, T. R., Bouvier, J., Moraux, E., 
    James, D. J., \& M\'enard, F. 2005, \aap, 434, 939
\bibitem[Kroupa(2002)]{kro02} Kroupa, P. 2002, Science, 295, 82
\bibitem[Lada(2006)]{cjl06} Lada, C. J. 2006, \apj, 640, L63
\bibitem[Lada et al.(1978)]{lect78} Lada, C. J., Elmegreen, B. G., Cong, H.-I.,
    \& Thaddeus, P. 1978, \apj, 226, L39
\bibitem[Lee \& Sung(1995)]{ls95} Lee, S.-W., \& Sung, H. 1995, JKAS, 28, 45
\bibitem[Luhman et al.(2003)]{kll03} Luhman, K. L., Brice\~no, C., Stauffer,
    J. R., \& Hartmann, L. 2003, \apj, 590, 348
\bibitem[Metchev \& Hillenbrand(2009)]{mh09} Metchev, S. A., \& Hillenbrand, L.
    A. 2009, \apjs, 181, 62
\bibitem[Moraux et al.(2003)]{mbsc03} Moraux, E., Bouvier, J., Stauffer, J. R.,
    \& Cuillandre, J.-C. 2003, \aap, 400, 891
\bibitem[Motto et al.(1998)]{man98} Motto, F., Andre, P., \& Neri, R. 1998,
    \aap, 336, 150
\bibitem[Muench et al.(2002)]{mlla02} Muench, A. A., Lada, E. A., Lada, C. J., 
    \& Alves, J. 2002, \apj, 573, 366
\bibitem[Oliver et al.(1996)]{omt96} Oliver, R. J., Masheder, M. R. W., \&
    Thaddeus, P. 1996, \aap, 315, 578
\bibitem[Padoan \& Nordlund(2002)]{pn02} Padoan, P., \& Nordlund, A. 2002, 
    \apj, 576, 870
\bibitem[Palla \& Stahler(2000)]{ps00} Palla, F., \& Stahler, S. W. 2000,
    \apj, 525, 772
\bibitem[Park et al.(2000)]{psbk00} Park, B.-G., Sung, H., Bessell, M. S., \&
    Kang, Y. H. 2000, \aj, 120, 894
\bibitem[Povich \& Whitney(2010)]{pw10} Povich, M. S., \& Whitney, B. A.
    2010, \apjl, in press (arXiv1004.1712)
\bibitem[Preibisch \& Zinnecker(1999)]{pz99} Preibisch, T., \& Zinnecker, H.
    1999, \aj, 117, 2381
\bibitem[Ram\'irez et al.(2004)]{svr04} Ram\'irez, S. V. et al. 2004, \aj,
    127, 2659
\bibitem[Rebull et al.(2002)]{lmr02} Rebull, L. M. et al. 2002, \aj, 123, 1528
\bibitem[Rieke et al.(2004)]{mips} Rieke et al. 2004, \apjs, 154, 25
\bibitem[Robitaille et al.(2006)]{tpr06} Robitaille, T. P., Whitney, B. A.,
    Indebetouw, R., Wood, K., \& Denzmore, P. 2006, \apjs, 167, 256
\bibitem[Robitaille et al.(2007)]{tpr07} Robitaille, T. P., Whitney, B. A.,
    Indebetouw, R., \& Wood, K. 2007, \apjs, 169, 328
\bibitem[Salpeter(1955)]{ees55} Salpeter, E. E. 1955, \apj, 121, 161
\bibitem[Scalo(1986)]{jms86} Scalo, J. M. 1986, Fundam. Cosmic Phys., 11, 1 
\bibitem[Scalo(2005)]{sca05} Scalo, J. M. 2005, in The Initial Mass Function
    50 Years Later, ed. E. Corbelli, F. Palla, \& H. Zinnecker (Dordrecht: 
    Springer), p.23
\bibitem[Schmidt-Kaler(1982)]{lb82} Schmidt-Kaler, K. in Landolt-B\"orenstein,
    Vol. 2b, p19, p453
\bibitem[Schnurr et al.(2008)]{scc08} Schnurr, O., Casoli, J., Chen\'e, A.-N., 
    Moffat, A. F. J., \& St-Louis, N. 2008, \mnras, 389, L38
\bibitem[Schwartz(1987)]{prs87} Schwartz, P. R. 1987, \apj, 320, 258
\bibitem[Shu et al.(2004)]{sla04} Shu, F., Li, Z.-Y., \& Allen, A. 2004, \apj,
    601, 930
\bibitem[Siess et al.(2000)]{sdf00} Siess, L., Dufour, E., \& Forestini, M.
    2000, \aap, 358, 593
\bibitem[Skrutskie et al.(2006)]{2mass} Skrutskie, M. F. et al. 2006, \aj,
    131, 1163
\bibitem[Smith et al.(2009)]{scb09} Smith, R. J., Clark, P. C., \& Bonnell, 
    I. A. 2009, \mnras, 396, 830
\bibitem[Sterzik \& Durisen(2003)]{sd03} Sterzik, M. F. \& Durisen, R. H. 
    2003, \aap, 400, 1031
\bibitem[Sung et al.(2004)]{sbc04} Sung, H., Bessell, M. S., \& Chun, M.-Y.
    2004, \aj, 128, 1684 (Paper I)
\bibitem[Sung et al.(2005)]{sbc05} Sung, H., Bessell, M. S., \& Chun, M.-Y.
    2005, in (ASP Conf. Ser. 362, The seventh Pacific Rim Conference on 
    Stellar Astrophysics, eds. Y. W. Kang, H.-W. Lee, and K.-C. Leung 
    (San Francisco, CA; ASP) 275 (Paper II)
\bibitem[Sung et al.(2008)]{sbc08} Sung, H., Bessell, M. S., Chun, M.-Y.,
    Karimov, R., \& Ibrahimov, M. 2008, \aj, 135, 441 (Paper III)
\bibitem[Sung et al.(1997)]{sbl97} Sung, H., Bessell, M. S., \& Lee, S.-W.
    1997, \aj, 114, 2644 (SBL97)
\bibitem[Sung et al.(1998)]{sbl98} Sung, H., Bessell, M. S., \& Lee, S.-W.
    1998, \aj, 115, 734
\bibitem[Sung et al.(2000)]{scb00} Sung, H., Chun, M.-Y., \& Bessell, M. S.
    2000, \aj, 120, 333
\bibitem[Sung et al.(2009)]{ssb09} Sung, H., Stauffer, J. R., \& Bessell, M. S.
    2009, \aj, 138, 1116
\bibitem[Tauber et al.(1993)]{tlg93} Tauber, J. A., Lis, D. C., \& Goldsmith,
    P. F. 1993, \apj, 403, 202
\bibitem[Teixeira et al.(2006)]{pst06} Teixeira, P. S. et al. 2006, \apj, 636,
    L45
\bibitem[Teixeira et al.(2007)]{tzl07} Teixeira, P. S., Zapata, L. A., \&
    Lada, C. J. 2007, \apj, 667, L179
\bibitem[Wright et al.(2010)]{wddv10} Wright, N. J., Drake, J. J., Drew, J. E.,
    \& Vink, J. S. 2010, arXiv:1003.2463
\bibitem[Zinnecker(1984)]{hz84} Zinnecker, H. 1984, \mnras, 210, 43
\end{thebibliography}
\end{document}